\documentclass[11pt]{article}
\usepackage{jcappub,natbib,float,caption,comment,bm}

\def\be{\begin{equation}}
\def\ee{\end{equation}}
\def\ba#1\ea{\begin{align}#1\end{align}}
\newcommand{\vs}{\nonumber\\}

\renewcommand{\v}[1]{\mathbf{#1}}
\newcommand{\vx}{\v{x}}
\newcommand{\vr}{\v{r}}
\newcommand{\vk}{\v{k}}

\newcommand{\refeq}[1]{Eq.~(\ref{eq:#1})}          
\newcommand{\refeqs}[2]{Eqs.~(\ref{eq:#1})--(\ref{eq:#2})}          
\newcommand{\reffig}[1]{figure~\ref{fig:#1}}          
\newcommand{\refFig}[1]{Figure~\ref{fig:#1}}          
\newcommand{\refsec}[1]{section~\ref{sec:#1}}          
\newcommand{\refapp}[1]{Appendix~\ref{app:#1}}

\newcommand{\Om}{\Omega_m}

\newcommand{\rhob}{\bar\rho}

\renewcommand{\d}{\delta}
\newcommand{\tOm}{\tilde{\Omega}_m}
\newcommand{\tOK}{\tilde{\Omega}_K}

\def\iMpch{\,h\,{\rm Mpc}^{-1}}

\newcommand{\eps}{\epsilon}
\newcommand{\D}{\Delta}
\def\O{\mathcal{O}}

\newcommand{\<}{\left\langle}
\renewcommand{\>}{\right\rangle}

\title{The angle-averaged squeezed limit of nonlinear matter $\bm{N}$-point functions}
\author[a]{Christian Wagner,}
\author[a]{Fabian Schmidt,}
\author[a]{Chi-Ting Chiang}
\author[a,b]{and Eiichiro Komatsu}
\affiliation[a]{Max-Planck-Institut f\"ur Astrophysik, Karl-Schwarzschild-Str. 1, 85741 Garching, Germany}
\affiliation[b]{Kavli Institute for the Physics and Mathematics of the
Universe, Todai Institutes for Advanced Study, the University of Tokyo,
Kashiwa, Japan 277-8583 (Kavli IPMU, WPI)}
\emailAdd{cwagner@mpa-garching.mpg.de}
\abstract{
We show that in a certain, angle-averaged squeezed limit, the 
$N$-point function of matter is related to the response of the matter 
power spectrum to a long-wavelength density perturbation,
$P^{-1}d^nP(k|\delta_L)/d\delta_L^n|_{\delta_L=0}$, with $n=N-2$.  
By performing N-body simulations with a homogeneous overdensity
superimposed on a flat Friedmann-Robertson-Lema\^itre-Walker (FRLW) universe
using the \emph{separate universe} approach, we 
obtain measurements of the nonlinear matter power spectrum response up to
$n=3$, which is equivalent to measuring the fully nonlinear matter
$3-$ to $5-$point function in this squeezed limit.  
The sub-percent to few percent accuracy of those measurements is unprecedented.
We then test the hypothesis that nonlinear $N$-point functions at a given
time are a function of the linear power spectrum at that time, 
which is predicted by standard perturbation
theory (SPT) and its variants that are based on the ideal pressureless fluid
equations. Specifically, we compare the responses computed from the
separate universe simulations and simulations with a rescaled initial (linear) power spectrum amplitude.   
We find discrepancies of 10\% at $k\simeq 0.2 - 0.5 \iMpch$ for $5-$ to
$3-$point functions at $z=0$. The discrepancy occurs at higher wavenumbers at $z=2$.
Thus, SPT and its variants, carried out to arbitrarily high order, are guaranteed to fail to describe matter $N$-point functions ($N>2$) around that scale.
}
\begin{document}
\maketitle
\flushbottom

%%%%%%%%%%%%%%%%%%%%%%%%%%%%%%%%%%%%%%%%%%%%%%%%%%%%%%%%%%%%%%%%%%%%%%%%%%%%%
%%%%%%%%%%%%%%%%%%%%%%%%%%%%%%%%%%%%%%%%%%%%%%%%%%%%%%%%%%%%%%%%%%%%%%%%%%%%%
\section{Introduction}
\label{sec:intro}

Mode-coupling plays a fundamental role in cosmology. Long and short
wavelength modes are coupled by nonlinear gravitational evolution of
matter density fluctuations and the formation of dark matter halos and
galaxies, as well as via the physics of inflation, i.e., primordial
non-Gaussianity.  
The traditional way of characterizing the mode coupling
is to study $N$-point correlation functions of matter, halo, and galaxy
density fields, as well as of fluctuations of the cosmic microwave background,  where the mode coupling enters at $N > 2$.  
The $N$-point functions of matter density fields in particular contain a rich amount
of information on gravity and the expansion history of the Universe.  At
$N > 2$, gravity produces specific non-Gaussian signatures that can be used to
test the physics of structure formation in the Universe, and must be
accounted for when attempting to connect large-scale structure with the
statistics of the initial conditions in order to search for primordial
non-Gaussianity.

A major challenge is that, beyond the power spectrum, the calculation of
(connected) nonlinear 
$N$-point functions becomes increasingly difficult.  Consider the 
gold standard for predicting the statistics of matter density, N-body simulations.  First, a large number
of modes is necessary to measure the correlation functions with
sufficient precision since the
cosmic variance noise is significant. This demands large computational
resources.  Second, higher $N$-point functions
are more sensitive to transients \cite{crocce/pueblas/scoccimarro:2006,mccullagh/jeong} 
from the finite starting redshift
and to mass resolution effects.  Third, estimators for higher
$N$-point functions become more computationally intensive and difficult 
to handle.  On the theoretical side, the predictions for the bispectrum
and higher $N$-point functions likewise become more cumbersome, 
with terms at any given order in perturbation theory rapidly proliferating
with $N$.  An analogous effect happens in the halo model, with 1- to $N$-halo terms
having to be calculated for the $N$-point function.  

%!!!!!!!!!!!!!!!!!!!!!!!!!!!!!!!!!!!!!!!!!!!
\begin{figure}[t]
\centering
\includegraphics[width=0.7\textwidth, trim=0cm 19cm 0cm 2cm]{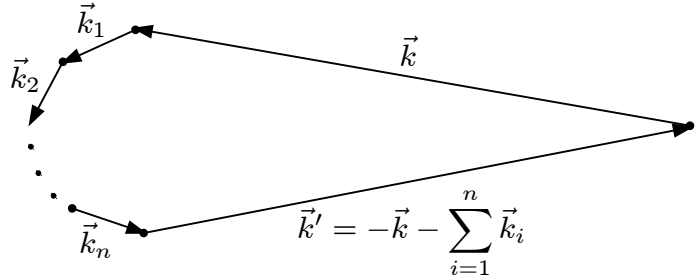}
\caption{Sketch of the squeezed limit configuration of matter $N$-point functions considered in this paper. $\vk_1,\cdots,\vk_n$ denote the long-wavelength modes which are spherically averaged in \refeq{Sdef}, while $\vk,\,\vk'$ denote
the small-scale modes which are allowed to be fully nonlinear.}
\label{fig:sketch}
\end{figure}
%!!!!!!!!!!!!!!!!!!!!!!!!!!!!!!!!!!!!!!!!!!!

This provides the motivation to study a certain limit considered in this
paper where the nonlinear $N$-point functions become simpler and
physically more transparent.
The limit we consider is a specific case of the so-called ``squeezed
limit'', where there 
is a hierarchy between two large wavenumbers $\vk,\,\vk'$ and $N-2$
small wavenumbers $\vk_1,\cdots \vk_n$. The configuration corresponding to this limit is illustrated in \reffig{sketch}.  

The squeezed limit of dark matter
$N$-point functions has recently
been the subject of a large body of work in the context of the so-called
``consistency relations'' \cite{r7,r8,r6,r4,r5,r2,valageas1,valageas2,r1,r3,nishimichi/valageas:2014,ben-dayan,r9}.  
The contributions to $N$-point functions in the squeezed limit are ordered
by the ratio of wavenumbers $k_i/k$, which is assumed to be much less than
one.  The lowest order contributions, up $\propto (k_i/k)^{-1}$ when the
$N$-point function is written in terms of the overdensity $\d$, 
are fixed by the requirement that a uniform potential perturbation 
as well as a uniform velocity (boost) do not lead to any locally
observable effect on the density field, as demanded by the 
equivalence principle \cite{r7,r8,valageas1,CFCpaper2}.  They are
also referred to as ``kinematical contributions''.  Here we focus
on the next order contribution, $\propto (k_i/k)^0$, which is the
lowest order at which a \emph{physical} coupling of long- and
short-wavelength modes happens.  More precisely, the contributions
at this order correspond to the impact of a uniform long-wavelength
density or tidal perturbation.  When considering equal-time $N$-point functions, which we do throughout, and subhorizon perturbations $k_i \gg aH$, the kinematical contributions disappear, and the physical $(k_i/k)^0$ contributions are 
the leading contribution to the $N$-point function in the squeezed limit.\footnote{This extends to $k_i \lesssim aH$ if the density perturbation is written in synchronous-comoving gauge \cite{CFCpaper2}.}

In this paper, we disregard
tidal fields, which leads us to first angle-average over the $N-2$ small momenta (wavenumbers) in the $N$-point function.  
Specifically, we consider $\mathcal{S}_{N-2}$ defined through
\ba
\mathcal{S}_{N-2}(k, k'; k_1,\cdots,k_{N-2}) \equiv\:& \int \frac{d^2\hat{\vk}_1}{4\pi} \cdots
\int \frac{d^2\hat{\vk}_{N-2}}{4\pi}
\< \d(\vk) \d(\vk') \d(\vk_1) \cdots \d(\vk_{N-2}) \>'_c \,,
\label{eq:Sdef}
\ea
where $\hat{\vk}_i$ are unit vectors and $\< \d(\vk_1) \cdots \d(\vk_{N}) \>'_c$ denotes the nonlinear connected matter 
$N$-point function with the  momentum constraint
$(2\pi)^3\delta_D(\vk_1+\cdots+\vk_N)$ dropped.  Note that the momentum
constraint fixes $\vk'$ in terms of $\vk$ and $\vk_1,\dots,\vk_{N-2}$.  
We now let $k_1,\dots, k_{N-2}$ go to zero, and normalize the result by
the nonlinear power spectrum $P(k)$ and the
linear power spectra
$P_l(k_1)\cdots P_l(k_{N-2})$ to obtain a dimensionless quantity:
\be
R_{N-2}(k) = \lim_{k_i\to 0}\; \frac{\mathcal{S}_{N-2}(k,k'; k_1,\cdots k_{N-2})}{P(k) P_l(k_1) \cdots P_l(k_{N-2})}\,.
\label{eq:Rnpoint}
\ee
Note that in this limit, spatial homogeneity enforces $\vk' = -\vk + \O(k_i/k)$, so
that (for statistically isotropic initial conditions) the r.h.s. only
depends on $k$.   
In \refapp{sqlimit} (see also \cite{valageas2}), we show that the $R_n(k)$ \emph{exactly correspond to 
the power spectrum response functions}, which quantify the change
in the nonlinear matter power spectrum to an infinite-wavelength density
perturbation.  These response functions are defined as the coefficients 
of the expansion of the power spectrum in the
\emph{linearly extrapolated initial overdensity} $\d_{L0}$:
\be
P(k, t|\d_{L0}) = \sum_{n=0}^\infty \frac1{n!} R_n(k,t) \left[\d_{L0} \hat D(t)\right]^n\: P(k, t)\,,
\label{eq:Pkexpansion}
\ee
where $P(k, t|\d_{L0})$ is the nonlinear matter power spectrum at time $t$
in the presence of a homogeneous (infinite-wavelength) density perturbation, and
$\hat D(t)$ is the linear growth factor normalized to unity today.  We
have set $R_0(k,t) = 1$ by definition.  Thus, by measuring $R_n$, we measure
the angle-averaged squeezed limit (\refeq{Rnpoint}) of the nonlinear matter
$(n+2)$-point function.  

For $n=1$, the response $R_1$ describes the angle-averaged squeezed
limit bispectrum.  This relation has been
derived several times in the literature (e.g.,
\cite{sherwin/zaldarriaga:2012,r1,ben-dayan}).
Refs.~\cite{takada/hu:2013,li/hu/takada:2014} considered the case of $n=1$
in the context of the power spectrum covariance.  
Ref.~\cite{valageas2} considered
the general angle-averaged case of $n$ long modes and $l$ short modes;  
however, a hierarchy between all of the long modes $k_i \ll k_{i+1}$ was 
assumed which we do not assume here.  

Independently of the derivation of Eqs.~\eqref{eq:Rnpoint} and \eqref{eq:Pkexpansion}, we
here present accurate measurements of $R_n$ for $n=1, 2, 3$ using N-body
simulations which do not rely on approximations. Specifically,
we resort to N-body simulations with
an external homogeneous overdensity imposed via the \emph{separate universe} approach described in Ref.~\cite{sep1} (see also \cite{mcdonald:2003,sirko:2005,li/hu/takada:2014} when the overdensity can be approximated to be small).  
A flat FLRW universe with a homogeneous overdensity is exactly equivalent
to a different, curved FLRW universe \cite{baldauf/etal:2011,CFCpaper2}, so that N-body simulations in
this modified cosmology provide, in principle, the exact result for
the response functions $R_n(k)$. This in turn corresponds to the exact (in the limit of infinite volume and resolution) measurement of the squeezed-limit $N$-point function (\refeq{Rnpoint}). Ref.~\cite{li/hu/takada:2014} presented
simulation measurements of $R_1(k)$. 
In Ref.~\cite{sep1}, in which we introduced the separate universe simulation technique, we already briefly presented 
a subset of the results shown here (the so-called ``growth-only response'' defined later in
this paper) for $n=2$ and 3 as a sample application of the method. 
Here, we present the full results for $n=2$ and 3 for the first time.

Many semi-analytical approaches to nonlinear large-scale structure 
assume that nonlinear matter statistics can be described as a unique function of the
linear matter power spectrum, i.e. the power spectrum of initial fluctuations
linearly extrapolated to a given time.  
In the context of consistency relations, this approximation has been
studied in, e.g., Refs.~\cite{valageas2,r1}.  This ansatz is motivated by the
fact that in Einstein-de Sitter (flat matter-dominated universe), and to a very good approximation
in $\Lambda$CDM, the perturbation theory predictions factorize into
powers of the linear growth factor and convolutions of products of the
initial matter power spectra and time-independent functions. Another way to phrase this ansatz is that
nonlinear large-scale structure only depends on the normalization of the
fluctuations at a given time, and not on the growth \emph{history}.  
In the context of squeezed-limit $N$-point functions, this ansatz can be 
tested quantitatively by comparing the outputs of separate universe
simulations at a given time with
simulations in which the initial amplitude of fluctuations is rescaled
to match the \emph{linear} power spectrum at the same time.  
The difference between these ``rescaled initial amplitude'' simulations
and the separate universe simulations corresponds to the error made in
the ansatz of assuming that the linear power spectrum at a given time
uniquely describes nonlinear large-scale structure at the same
time.  Ref.~\cite{li/hu/takada/3} studied this for $n=1$ and found that the two
simulations differ in the nonlinear regime.  
Ref.~\cite{nishimichi/valageas:2014} performed a closely related test using
the matter bispectrum.  In this paper, we study this comparison
in more detail and for $n=1, 2$ and 3.

The outline of the paper is as follows. We develop semi-analytic 
predictions for the power spectrum response in \refsec{response}.  
We then describe the N-body simulations used in this paper in \refsec{simsec}. 
Results and comparisons are presented in \refsec{results}, and we 
conclude in \refsec{conclusion}.  
The appendices present the proof of eqs.~\eqref{eq:Rnpoint} and \eqref{eq:Pkexpansion}
(\refapp{sqlimit}) as well as various useful results on the separate
universe picture necessary for the analytical approaches in \refsec{response}.

%%%%%%%%%%%%%%%%%%%%%%%%%%%%%%%%%%%%%%%%%%%%%%%%%%%%%%%%%%%%%%%%%%%%%%%%%%%%%
%%%%%%%%%%%%%%%%%%%%%%%%%%%%%%%%%%%%%%%%%%%%%%%%%%%%%%%%%%%%%%%%%%%%%%%%%%%%%
\section{Power spectrum response}
\label{sec:response}
We define the $n$th-order response function $R_n(k)$ of the power
spectrum as the $n$th derivative of the power spectrum with respect to
the linearly extrapolated (or Lagrangian) overdensity $\delta_L$,
normalized by the power spectrum. The definition consistent with
\refeq{Pkexpansion} is
\be
R_n(k,t)=\frac{1}{P(k)}\frac{d^n P(k,t|\delta_L)}{d(\delta_{L}(t))^n}\bigg|_{\delta_L=0}\,,
\label{eq:def_response}
\ee
where $\delta_L(t) \equiv \delta_{L0} \hat D(t)$.  In the following, we will
frequently suppress the time argument for clarity.  
Analogously, one can define the power spectrum response functions with
respect to the fully evolved (or Eulerian) nonlinear overdensity $\delta_\rho$. 
Since we can expand the nonlinear overdensity in powers of $\delta_{L}$ with known coefficients via the spherical collapse (see \refapp{growthMD}), the $n$th-order Eulerian response function is given by a sum of $R_m$ with $m\le n$.  
In this paper, motivated by the relation \refeq{Rnpoint}, we mainly consider the Lagrangian response functions.  
In the remainder of this section, we develop semi-analytic models for the 
response functions based on the separate universe picture.  

%%%%%%%%%%%%%%%%%%%%%%%%%%%%%%%%%%%%%%%%%%%%%%%%%%%%%%%%%%%%%%%%%%%%%%%%%%%%%
\subsection{Separate universe picture}
\label{sec:responsegen}

An infinite-wavelength adiabatic density perturbation $\d_\rho$
behaves like an independent curved ``separate'' universe
\cite{lemaitre:1933,barrow/saich:1993,mcdonald:2003,sirko:2005,baldauf/etal:2011},
in which $\delta_\rho$ is absorbed in a modified background matter density
by a modification of the cosmological
parameters.  A positive overdensity implies a slower expansion, which can
be quantified by the relative difference in the scale factor of the
background and modified cosmology $\delta_a(t)=\tilde a(t)/a(t)-1$,
where here and throughout the paper a tilde denotes quantities in the
modified, separate universe cosmology.  On the other hand, $a(t)$ refers
to the fiducial cosmology, for which we aim to calculate the response.  
Consequently, the comoving coordinates of the two cosmologies are related by
\be
 \vx = \frac{\tilde a(t)}{a(t)}\, \tilde \vx 
= [1 + \d_a(t)] \tilde \vx\,.
\label{eq:xrescaleNL}
\ee
Furthermore, due to mass conservation, the fractional difference in the scale factor is related to the overdensity $\delta_\rho$ by
\be
1 + \d_\rho(t) = [1 + \d_a(t)]^{-3}\,.
\label{eq:dadrho}
\ee
Using the separate universe picture, we can regard the matter power spectrum in this patch just as that of a region with no homogeneous overdensity but properly modified cosmology. The modification of the cosmology is such that the shape of the linear power spectrum is unchanged, since the ratio of photon, baryon, and cold dark matter densities is unmodified; moreover, the transfer function parameters are unchanged: $\tilde \Omega_m \tilde h^2=\Omega_m  h^2$ and $\tilde \Omega_b \tilde h^2=\Omega_b  h^2$.  Thus, only the growth of structure is affected.

The power spectrum that enters in the response given by
\refeq{def_response} is defined with respect to the background density
and comoving coordinates of the fiducial cosmology.  Hence, the power
spectrum calculated for the modified cosmology has
to be mapped to that with respect to the background density and comoving
coordinates of the fiducial 
cosmology. This mapping, described in more detail below, yields the ``reference
density'' and ``dilation'' contributions to the response \cite{sherwin/zaldarriaga:2012,r1,ben-dayan,li/hu/takada/3}.  
These can be calculated exactly at any scale $k$ to any given order given
the nonlinear matter power spectrum in the \emph{fiducial} cosmology.  
That is, we do not need to run separate simulations to calculate these effects.   They are thus merely ``projection effects'', unlike the effect of the modified cosmology on the growth of structure, which requires a simulation in order to provide an accurate estimate.  
Let us denote the power spectrum for the modified cosmology as $\tilde P(\tilde k)$.  Then, the reference density effect simply 
rescales the power spectrum,
\be
P(k) \stackrel{\rm ref.~density}{=} \left[1+\delta_\rho\right]^2 \tilde P(k)\,,
\ee
where the argument of $\tilde{P}(k)$ is not modified. 
The dilation effect due to the change in the coordinates given by
\refeq{xrescaleNL} implies $k\to \tilde k = (1+\d_a) k$ and changes the power spectrum by (see \refapp{xrescalePk})
\be
P(k) \stackrel{\rm dilation}{=} \left[1 + \d_a\right]^3 \tilde P([1 + \d_a] k) \,.
\ee
Putting the two together and using \refeq{dadrho} yields
\be
P(k) = \left[1+\delta_\rho\right] \tilde P\left([1 + \d_a] k\right) \,, 
\label{eq:refdensity_dilation}
\ee
where all quantities are evaluated at some fixed time $t$.  
Note that one prefactor of $1+\d_\rho$ cancels, since the
effect of the increased density is partially canceled by the corresponding
decrease in physical volume.  
For a flat matter-dominated fiducial cosmology, it is straightforward to derive series solutions for $\delta_\rho$ and $\delta_a$ (\refapp{growthMD}) of the form
\ba
\delta_a(t) =\:& \sum_{n=1}^\infty e_n \left[\d_{L0} \hat D(t) \right]^n \vs
\delta_\rho(t) =\:& \sum_{n=1}^\infty f_n \left[\d_{L0} \hat D(t) \right]^n\,,
\label{eq:seriessol}
\ea
where $\hat D(t) = D(t)/D(t_0)$ is the fiducial growth factor normalized to one at the epoch $t_0$ to which we extrapolate $\d_{L0} = \delta_L(t_0)$, and
$e_n,\, f_n$ are rational numbers.

The third contribution to $R_n$ comes from the effect of the modified
cosmology on the growth of structure, which as mentioned above is the
physical contribution which requires N-body simulations for an accurate measurement.  
We thus define a set of \emph{growth-only response functions} $G_n(k)$ which
isolate the nontrivial effect of the long-wavelength perturbation on the growth of small-scale structure,
\be
G_n(k) \equiv \frac{1}{P(k)}\frac{d^n \tilde P(k)}{d\delta_{L}^n}\bigg|_{\delta_L=0}\,.
\label{eq:def_Gn}
\ee
That is, $G_n$ are defined as $R_n$ without the contributions from the
reference density and dilation given by \refeq{refdensity_dilation}.  
This definition is an extension of the similar decomposition for $n=1$ shown in 
refs.~\cite{sherwin/zaldarriaga:2012,r1,ben-dayan,li/hu/takada/3}.  
Thus, the formula for the power spectrum (w.r.t. global coordinates) in
the presence of a long-wavelength overdensity is given by
\be
P(k|\d_L) = \left[1+\delta_\rho\right] \left[\left(1 + \sum_{n=1}^\infty \frac1{n!} G_n(\tilde k) \delta_L^n \right)P(\tilde k)\right]_{\tilde k = [1 + \d_a] k} \,.
\label{eq:PkdL}
\ee
Clearly, by the Leibniz rule, at any given order $n$ the total or ``full'' response $R_n(k)$ [\refeq{def_response}] is composed of the functions $G_m(k)$ and the numbers $e_m,\,f_m$ with $1 \leq m \leq n$, where the $e_m$ multiply derivatives of $G_l(k)$  and  $P(k)$ with respect to $k$ (up to the $n$-th derivative).  Specifically, the first three full response functions are given by
\ba 
R_1(k) =\:&  f_1+e_1  \frac{k P'(k)}{P(k)} 	+ G_1(k)\,,  \label{eq:R1} \\
\frac{R_2(k)}{2} =\:&  f_2+e_2 \frac{k P'(k)}{P(k)}  +e_1^2 \frac{k^2 P''(k)}{2P(k)}  
		+\frac{G_2(k)}{2} +f_1 e_1 \frac{k P'(k)}{P(k)} \vs 
		&+f_1 G_1(k)
		+e_1 \frac{k P'(k)}{P(k)} G_1(k) 
		+e_1 k G_1'(k)\,,  \label{eq:R2}  		\\
\frac{R_3(k)}{6} =\:& 		f_1 G_1(k) e_1  \frac{k P'(k)}{P(k)} 
		+f_3+\frac{G_3(k)}{6}+e_3 \frac{k P'(k)}{P(k)} 
		+f_1 \frac{G_2(k)}{2} +f_1 e_2 \frac{k P'(k)}{P(k)} \vs 
		& +f_1 e_1^2 \frac{k^2 P''(k)}{2P(k)}  
		+f_2 G_1(k) +f_2 e_1 \frac{k P'(k)}{P(k)} 
		+(f_1 e_1+e_2) k G_1'(k)
		+e_1^2  \frac{k^2 G_1''(k)}{2}\vs 
		&+e_1 k \frac{G_2'(k) }{2}
		+e_1^2 \frac{k P'(k)}{P(k)}  k G_1'(k) 
		+e_1^3 \frac{k^3P'''(k)}{6 P(k)}+ 2 e_1 e_2 \frac{k^2 P''(k)}{2P(k)}  \vs
		&+e_1 \frac{k P'(k)}{P(k)}  \frac{G_2(k)}{2}+G_1(k) \left(e_2 \frac{k P'(k)}{P(k)} +e_1^2 \frac{k^2 P''(k)}{2P(k)} \right)\,, \label{eq:R3}  
\ea 
where the primes denote derivatives with respect to $k$. 

%%%%%%%%%%%%%%%%%%%%%%%%%%%%%%%%%%%%%%%%%%%%%%%%%%%%%%%%%%%%%%%%%%%%%%%%
\subsection{Linear power spectrum predictions}

We now evaluate Eq.~\eqref{eq:PkdL} for the simplest case, i.e., the response of the linear matter power spectrum.
In linear theory, the growth is scale-independent and given by the linear growth factor.  Thus, the growth-only response functions are scale-independent and just described by the linear growth factor in the modified cosmology $\tilde D(t)$,
\be
G_n^{\rm linear} = \frac{1}{D^2}\frac{d^n (\tilde D^2)}{d\delta_{L}^n}\bigg|_{\delta_L=0}\,.
\label{eq:Gn_linear}
\ee
A perturbative expansion of $\tilde D$ in powers of $\delta_L$ for a flat matter-dominated fiducial cosmology is derived in \refapp{smallgrowth}, with the result given in \refeq{smallgrowth},
\be
\tilde D(t) = D(t)\left\{1 + \sum_{n=1}^\infty g_n \left[\d_{L0} \hat D(t)\right]^n \right\}\,.
\label{eq:Dtilde}
\ee
Thus, for an Einstein-de Sitter fiducial universe (and to high accuracy in
$\Lambda$CDM), the linear response functions are simply constants.  
Inserting the result from \refeq{gn}, we obtain
\ba
\Big\{ G_n^{\rm linear} \Big\}_{n=1,\cdots 4} =\:& 
\left\{
\frac{26}{21},\;
\frac{3002}{1323},\;
\frac{240272}{43659},\;
\frac{197919160}{11918907}
\right\}\,.
\label{eq:derivsGO}
\ea
\refeq{PkdL} evaluated for the linear matter power spectrum $P_l(k,t)$ then becomes
\be
P_l(k, t | \d_L) = [1 + \d_\rho(t)] \left(\frac{\tilde D(t)}{D(t)}\right)^2 P_{l,\rm fid}([1 + \d_a(t)] k, t)\,.
\ee
Inserting the series expansions
derived in \refapp{clMD} and \refapp{smallgrowth}, we obtain
\ba
P_l(k, t | \d_{L0}) =\:& \left(1 + \sum_{n=1}^\infty f_n [\d_{L0} \hat D(t)]^n \right) 
\left(1 + \sum_{n=1}^\infty g_n \left[\d_{L0} \hat D(t) \right]^n \right)^2 \vs
& \times P_{l,\rm fid}\left(\left [1 + \sum_{n=1}^\infty e_n [\d_{L0} \hat D(t)]^n \right] k, t\right)\,.
\label{eq:Plinresponse}
\ea
\refeq{Plinresponse} allows for
a consistent expansion in $\d_{L0}$.  Specifically, 
$d^n P_l(k) / d\d_{L0}^n$ is given by the $n$-th order coefficient in this
expansion, multiplied by $n!$.

%%%%%%%%%%%%%%%%%%%%%%%%%%%%%%%%%%%%%%%%%%%%%%%%%%%%%%%%%%%%%%%%%%%%%%%%%%%%%
\subsection{Nonlinear power spectrum predictions}
\label{sec:NLgrowth}

Beyond the linear matter power spectrum, the growth coefficients $G_n$ will become scale-dependent functions $G_n(k)$.  Consider now what standard 
perturbation theory (SPT) predicts.  The power spectrum prediction
is given by a series
\be
P^{\rm SPT}(k) = P_l(k) + P^{1-\rm loop}(k) + P^{2-\rm loop}(k) + \cdots\,,
\ee
where $P^{n-\rm loop}$ scales as
$[P_l]^n$.  In an Einstein-de Sitter universe, one can show
(e.g., \cite{bernardeau/etal:2001}) that the time- and scale-dependence
of each order in perturbation theory factorizes, so that one can write
\be
P^{\rm SPT}(k,t) = \hat D^2(t) P_l(k,t_0) + \hat D^4(t) P^{1-\rm loop}(k, t_0)
+ \hat D^6 P^{2-\rm loop}(k, t_0) + \cdots\,,
\label{eq:PSPT_EdS}
\ee
where $P^{n-\rm loop}(k, t_0)$ is a convolution of $n$ factors of $P_l(k, t_0)$
with \emph{time-independent} coefficients.  While \refeq{PSPT_EdS} is
only strictly correct in Einstein-de Sitter, it is used very commonly
for $\Lambda$CDM as well, since departures from the exact result are
typically of order 1\% or less, and since it simplifies the calculation
significantly. Various variants of SPT, such as the renormalized
perturbation theory (RPT) \cite{rpt}, share the same property.

In the context of this paper, \refeq{PSPT_EdS} allows for a very simple
evaluation of the growth-only response: as discussed above, the shape of
the linear power spectrum in the modified cosmology is unchanged, and
hence $\tilde P^{\rm SPT}(\tilde k)$ can be simply evaluated by replacing
the fiducial $\hat D(t)$ in \refeq{PSPT_EdS} with the modified one, \refeq{Dtilde}.  This is equivalent to assuming that the entire late-time cosmology dependence of the nonlinear matter power spectrum enters through the linear growth factor \cite{valageas2,r1,ben-dayan}.  

Apart from the SPT calculation, we can also apply this approximation to any prescription that maps a given linear power spectrum to a nonlinear one.  
In particular, we will show results for \textsc{halofit} \cite{smith/etal:2002}.  
In this case, where the dependence on the linear growth factor is not explicit, we instead compute the derivative with respect to the normalization of the linear power spectrum,
\be
\frac{d}{d\tilde D} \rightarrow \frac{d\tilde \sigma_8 }{d\tilde D}\frac{d}{d\tilde \sigma_8}\,,
\label{eq:growth_approx}
\ee
which at the redshift considered yields the equivalent change of the linear matter power spectrum.  This leads to
\be 
D^n \frac{d^nP(k)}{d\tilde D^n}\rightarrow \sigma_8^n \frac{d^nP(k)}{d\tilde \sigma_8^n}\,.
\ee
We use a five-point stencil with a step size of $0.75\%$ in $\sigma_8$ to compute numerically the derivatives with respect to $\sigma_8$.  
In conjunction with the change of the linear growth factor \refeq{Dtilde}, this allows us to compute the growth-only response $G_n(k)$ for perturbation theory as well as fitting formulae of the nonlinear matter power spectrum.  

Further, we can test this prescription  to all orders in SPT
calculations, and independently of fitting functions, by performing
simulations with a rescaled
initial power spectrum.  This is the subject of \refsec{growth_history}.

%%%%%%%%%%%%%%%%%%%%%%%%%%%%%%%%%%%%%%%%%%%%%%%%%%%%%%%%%%%%%%%%%%%%%%%%%%%%%
\subsection{Halo model predictions}
\label{sec:halomodel}

In the halo model (see \cite{cooray/sheth} for a review), all matter is
assumed to be contained within halos with a certain distribution of mass
given by the mass function, and a certain density profile. Along with
the clustering properties of the halos, these quantities then determine
the statistics of the matter density field on all scales including the
nonlinear regime. $N$-point functions can be conveniently decomposed
into 1- through $N$-halo pieces. In the following, we will follow the
most common halo model approach and assume a linear local bias of the
halos.  This is the most popular choice in the literature, although it can
clearly be improved upon (and is not strictly consistent, as we will see).  

Adopting the notation of Ref.~\cite{takada/hu:2013}, the halo model power
spectrum, $P_{\rm HM}(k)$, is given by
\ba
P_{\rm HM}(k) =\:& P^{\rm 2h}(k) + P^{\rm 1h}(k) \label{eq:PkHM}\,,\\
P^{\rm 2h}(k) =\:& \left[I^1_1(k)\right]^2 P_l(k)\,, \vs
P^{\rm 1h}(k) =\:& I^0_2(k,k)\,,\nonumber
\ea
where
\be
I^n_m(k_1,\cdots k_m) \equiv \int d\ln M\:n(\ln M) \left(\frac{M}{\rhob}\right)^m \, b_n(M)\, u(M|k_1) \cdots u(M|k_m)\,,
\label{eq:Inmdef}
\ee
and $n(\ln M)$ is the mass function (comoving number density per interval
in log mass), $M$ is the halo mass, $b_n(M)$ is the $n$-th order local bias 
parameter, and $u(M|k)$ is the dimensionless Fourier transform of the halo 
density profile, for which we use the NFW profile \cite{NFW}. We normalize
$u$ so that $u(M|k\to 0)=1$.  The notation given in \refeq{Inmdef}
assumes $b_0\equiv1$. $u(M|k)$ depends on $M$ through the scale radius
$r_s$, which in turn is given through the mass-concentration
relation. All functions of $M$ in \refeq{Inmdef} are also functions of
$z$ although we have not shown this for clarity. In the following,
we adopt the Sheth-Tormen mass function \cite{sheth/tormen} with
the corresponding peak-background split bias, and the mass-concentration
relation of Ref.~\cite{bullock/etal}. The exact choice of the latter only has 
a small impact on the predictions which does not affect our conclusions.  
A dependence of halo profiles on the background cosmology can however
change the response functions on small scales $k\gtrsim 1 \iMpch$ 
(see \refsec{haloprof} below).  
The derivations of the halo model response given below generalize
the linear response calculations of \cite{takada/hu:2013,posdeppk} to
arbitrary nonlinear order.  

\subsubsection{Total halo model response}

We now derive how the power spectrum given in \refeq{PkHM} responds to a
homogeneous (infinitely long-wavelength) density perturbation $\d_L$. 
For this, we consider the 1-halo and 2-halo
terms separately. The key physical assumption we make is that halo profiles
in \emph{physical} coordinates are unchanged by $\d_L$. 
That is, halos at a given mass $M$ in the presence of $\d_L$
have the same scale radius $r_s$ and scale density $\rho(r_s)$ as in the
fiducial cosmology.  We will discuss this assumption in \refsec{haloprof}.  
Given this assumption, the density perturbation
$\d_L$ then mainly affects the linear power spectrum, which determines
the halo-halo clustering (2-halo term), and the abundance of halos at a
given mass. 

We begin with the 2-halo term.  The response of the linear power spectrum
was derived in \refeq{Plinresponse} in the previous section.  The expression for the 2-halo term in
\refeq{PkHM} is simply the convolution (in real space) of the halo
correlation function in the linear bias model with the halo density
profiles.  By assumption, the density profiles do 
not change, hence $I^1_1$ only changes through the bias $b_1(M)$ and the mass
function $n(\ln M)$. The bias $b_N(M)$ quantifies the $N$-th order response
of the mass function $n(\ln M)$ to $\d_L$ \cite{mo/white:1995,PBSpaper}:
\be
b_N(M) = \frac{1}{n(\ln M)} \frac{\partial^N n(\ln M)}{\partial \d_L^N}\Big|_0\,, \quad\mbox{so that}\quad
\frac{\partial^N n(\ln M)}{\partial\d_L^N}\Big|_0 = b_N(M) n(\ln M)\,.
\label{eq:bNdef}
\ee
Thus,
\ba
\frac{\partial^N}{\partial\d_L^N} I^1_1(k)\Big|_{\d_L=0} =\:& \int d\ln M \: \left(\frac{M}{\rhob}\right)
\frac{\partial^N}{\partial\d_L^N} \left[ b_1(M) n(\ln M) \right]\Big|_{\d_L=0} u(M|k) 
= I^{N+1}_1(k)\,.
\label{eq:dI11}
\ea
Note that in the large-scale limit, $k\to 0$, this vanishes for $N \geq 1$ 
by way of the halo model consistency relation
\be
\int d\ln M\: n(\ln M)\left(\frac{M}{\rhob}\right) b_N(M) = 
\left\{
\begin{array}{ll}
1, & N = 1\,, \\
0, & N > 1\,.
\end{array}\right.
\ee
For finite $k$ however, \refeq{dI11} does not vanish.  We thus have
\be
I^1_1(k, t | \d_{L0}) = \sum_{n=0}^\infty \frac1{n!} I_1^{n+1}(k, t) [\hat D(t) \d_{L0} ]^n\,.
\ee
Thus, the two-halo term in the presence of $\d_{L0}$ becomes
\ba
P^{\rm 2h}(k, t | \d_{L0}) =\:& \left(1 + \sum_{n=1}^\infty f_n [\d_{L0} \hat D(t)]^n \right) 
\left(1 + \sum_{n=1}^\infty g_n \left[\d_{L0} \hat D(t) \right]^n \right)^2 
\label{eq:P2hresponse}\\
& \times 
\left(\sum_{n=0}^\infty \frac1{n!} I_1^{n+1}(k, t) [\hat D(t) \d_{L0} ]^n\right)^2
P_{l,\rm fid}\left(\left [1 + \sum_{n=1}^\infty e_n [\d_{L0} \hat D(t)]^n \right] k, t\right)\,. \nonumber
\ea
Note that we recover the tree-level result given in \refeq{Plinresponse} in the
large-scale limit. Strictly speaking, this expression is not consistent,
since the term $I^2_1$ implies a non-zero $b_2$ while in \refeq{PkHM} we
have assumed a pure linear bias.  
Note that in \refeq{P2hresponse} the dilation effect only enters in the
\emph{linear}, not 2-halo, power spectrum.  
This is a consequence of our assumption that halo
profiles do not change due to the long-wavelength density perturbation.

We now turn to the one-halo term. Given our assumption about density profiles,
this term is much simpler. The only effect is the change in the mass function,
which through \refeq{bNdef} becomes
\be
\frac{\partial^N}{\partial\d_L^N} I^0_2(k, k) = I^N_2(k,k)\,.
\ee
We thus obtain
\ba
P^{\rm 1h}(k, t | \d_{L0}) = \sum_{n=0}^\infty \frac1{n!} I_2^{n}(k,k, t) [\hat D(t) \d_{L0} ]^n\,.
\label{eq:P1hresponse}
\ea
Putting everything together, we obtain
\ba
P^{\rm HM}(k, t | \d_{L0}) =\:& \left(1 + \sum_{n=1}^\infty f_n [\d_{L0} \hat D(t)]^n \right) 
\left(1 + \sum_{n=1}^\infty g_n \left[\d_{L0} \hat D(t) \right]^n \right)^2 \vs
& \times 
\left(\sum_{n=0}^\infty \frac1{n!} I_1^{n+1}(k, t) [\hat D(t) \d_{L0} ]^n\right)^2
P_{l,\rm fid}\left(\left [1 + \sum_{n=1}^\infty e_n [\d_{L0} \hat D(t)]^n \right] k, t\right) \vs
& + \sum_{n=0}^\infty \frac1{n!} I_2^{n}(k,k, t) [\hat D(t) \d_{L0} ]^n
\,.
\label{eq:PHMresponse}
\ea
The contribution $\propto I_1^{n+1}$ (for $n > 0$) is numerically much smaller than
the other terms (see also the discussion in section~4.2.4 of \cite{posdeppk}).  Since it is much smaller than the overall accuracy of the halo model description, we will neglect it in the following.  This yields
\ba
P^{\rm HM}(k, t | \d_{L0}) =\:& \left(1 + \sum_{n=1}^\infty f_n [\d_{L0} \hat D(t)]^n \right) 
\left(1 + \sum_{n=1}^\infty g_n \left[\d_{L0} \hat D(t) \right]^n \right)^2 \vs
& \times 
\left(I_1^1(k, t) \right)^2
P_{l,\rm fid}\left(\left [1 + \sum_{n=1}^\infty e_n [\d_{L0} \hat D(t)]^n \right] k, t\right) \vs
& + \sum_{n=0}^\infty \frac1{n!} I_2^{n}(k,k, t) [\hat D(t) \d_{L0} ]^n
\,.
\label{eq:PHMsimplified}
\ea
Explicitly, the first and second order full response functions are given by
\ba
R_1^{\rm HM}(k) =\:&
\left[ f_1 + 2g_1 + e_1 \frac{d\ln P_l(k,t)}{d\ln k} \right]
P^{\rm 2h}(k,t) + I_2^1(k,k,t) \vs
R_2^{\rm HM}(k) =\:&
\bigg[ 2 f_2 + 2 f_1 g_1 + (f_1 + 2g_1) e_1 \frac{d\ln P_l(k,t)}{d\ln k} 
+ 2 g_1^2 + 4 g_2 \vs
& + 2 e_2 \frac{d\ln P_l(k,t)}{d\ln k}  
+ e_1^2 \frac1{P}\frac{d^2 P_l(k,t)}{d(\ln k)^2 }  \bigg]
P^{\rm 2h}(k,t) + I_2^2(k,k,t)\,.
\ea

% % % % % % % % % % % % % % % % % % % % % % % % % % % % % % % % % % % % % 
\subsubsection{Growth-only response}
\label{sec:haloprof}

We also derive the growth-only response functions in the halo model approach.  
Since the halo profiles are assumed fixed in
physical coordinates, this means that 
we need to rescale the halo model terms, $I^n_m$, accordingly.  
Following our discussion in \refsec{responsegen}, we have $\tilde{k} = (1+\d_a) k$, where $\tilde k$ is the comoving wavenumber with respect to the modified cosmology.  
We then obtain
\be
I^n_m\Big|_{\rm growth~only}(\tilde k_1, \cdots \tilde k_m)
= I^n_m\Big|_{\rm physical}\left(\frac{k_1}{1+\d_a(t)}, \cdots \frac{k_m}{1+\d_a(t)}\right)\,.
\label{eq:InmGO}
\ee
Inserting this into \refeq{PHMresponse} and performing a series expansion of $\delta_a$ in $\d_L$ then allows us to derive the growth-only response functions $G^{\rm HM}_n(k)$.  
Note that the NFW profile we assume
is uniquely determined by the scale radius $r_s(M)$ for a halo of mass $M$,  
which enters the coefficients 
defined in \refeq{Inmdef} in the combination $k r_s(M)$.  Thus, it is
easily possible to include a dependence of the scale radius $r_s(M)$, or
equivalently the halo concentration, on the long-wavelength density
in a similar way.  We will leave this for future work.

Quantitatively, the main contribution of the rescaling \refeq{InmGO} 
is from the 1-halo term $\propto I_2^n(k,k)$, i.e. the term in the last line of \refeq{PHMsimplified}.  The rescaling of the other instances of $I^n_m$ only
changes the response at the sub-percent level and we will neglect them in
the following.  
We then obtain for the growth-only contribution to the halo model power
spectrum
\ba
P^{\rm HM}(k, t | \d_{L0}) \stackrel{\rm growth~only}{=}\:& 
\left(1 + \sum_{n=1}^\infty g_n \left[\d_{L0} \hat D(t) \right]^n \right)^2 
\Big\{I_1^1\left[k, \, t\right] \Big\}^2
P_{l,\rm fid}(k, t) \vs
& + \sum_{n=0}^\infty \frac1{n!} I_2^{n}\left[A(\d_{L0},t)\,k, A(\d_{L0},t)\,k, t\right]\: [\hat D(t) \d_{L0} ]^n
\,,
\label{eq:PHMgo}
\ea
where
\be
A(\d_{L0}, t) = 
 \left(1 + \sum_{n=1}^\infty e_n [\d_{L0} \hat D(t)]^n \right)^{-1}\,.
\ee
This completes the derivation of the halo model response functions.

%%%%%%%%%%%%%%%%%%%%%%%%%%%%%%%%%%%%%%%%%%%%%%%%%%%%%%%%%%%%%%%%%%%%%%%%%%%
%%%%%%%%%%%%%%%%%%%%%%%%%%%%%%%%%%%%%%%%%%%%%%%%%%%%%%%%%%%%%%%%%%%%%%%%%%%
\section{N-body simulations}
\label{sec:simsec}

Before describing the separate universe simulations in detail, we summarize features common to all.  
All simulations are gravity-only simulations and are carried out with \textsc{Gadget-2} \cite{springel:2005}. The starting redshift is $z=49$ and the initial displacement field is computed using second-order Lagrangian perturbation theory. For each simulation, the particle load is $512^3$.
For the fiducial cosmology ($\delta_{L0}=0$),  we choose a flat
$\Lambda$CDM cosmology with cosmological parameters consistent with the current observational constraints: $\Omega_m=0.27$, $h=0.7$, $\Omega_b h^2=0.023$, $n_s=0.95$, $\sigma_8=0.8$,
and a comoving box size of $500\,h^{-1}$Mpc.  

\label{sec:sims}
\subsection{Separate universe simulations}
Using the separate universe approach presented in Ref.~\cite{sep1}, we
simulate separate universes corresponding to the linearly-evolved
present-day overdensities of $\delta_{L0}=0$, $\pm0.01$, $\pm0.02$, $\pm0.05$, $\pm0.07$, $\pm0.1$, $\pm0.2$, $\pm0.5$, $\pm0.7$, and $\pm1$.  Then, for the separate universes, the Hubble constant and the curvature fraction vary between  $\tilde h$: 0.447 to 0.883 and  $\tilde \Omega_K$: $-2.45$ to 0.372, respectively. The physical densities $\tilde \Omega_m \tilde h^2$, $\tilde \Omega_\Lambda \tilde h^2$, and $\tilde \Omega_b\tilde h^2$ as well as $n_s$ and the amplitude of the primordial curvature power spectrum remain the same.  

The initial conditions are set up as described in Ref.~\cite{sep1}. For each overdensity $\delta_{L0}$, we run the same 16 realizations of the Gaussian initial density field. Hence, by comparing relative differences between different $\delta_{L0}$ values but the same realization, most of the sample variance cancels out. The 16 realizations allow us to estimate the residual statistical error.

Given a fixed box size for the fiducial cosmology, there are two
reasonable choices for the box sizes of the modified cosmologies. Either
we match the respective comoving box sizes, i.e. the box size is $500\,
\tilde h /h$ in units of $\tilde h^{-1}$Mpc comoving, or we choose the
box sizes such that their physical sizes coincide with that of the
fiducial simulation at one specific output time $t_{\rm out}$,
i.e. $500\,  \tilde h a(t_{\rm out}) / [h\tilde a(t_{\rm out})] $ in units of $\tilde h^{-1}$Mpc comoving, where $a$ and $\tilde a$ are the scale factors of the fiducial and modified cosmology, respectively. The former choice is adequate if we are interested in the power spectrum response functions at the same comoving wavenumber, i.e. without the ``dilation'' effect. By using the mean density of the separate universe cosmology as the reference density when computing the power spectrum, we are further removing the ``reference density'' effect and are left with the growth-only response. In Ref.~\cite{sep1}, we have run simulations with this choice of box-size-matching to measure the growth-only response functions. Here, we reproduce the simulations with a higher mass resolution and compare the results with the models presented in this paper in \refsec{growth-only}.

In order to measure the full response functions, we run simulations for which we match the physical box size. We focus on two different output times $t_{\rm out}$ corresponding to $z=0$ and $z=2$ in the fiducial cosmology. As the physical size can only be matched at one specific time, we have to run a new set of simulations for each output time. The results of these simulations are presented in \refsec{full}.

%%%%%%%%%%%%%%%%%%%%%%%%%%%%%%%%%%%%%%%%%%%%%%%%%%%%%%%%%%%%%%%%%%%%%%%%%%%
\subsection{Simulations with rescaled initial amplitude}
\label{sec:growth_history}

We also investigate how well the effect of a homogeneous overdensity on
the growth of structure can be modelled by a change in the amplitude of
the linear power spectrum. To this end, we additionally run a set of simulations for which we always assume the fiducial cosmology but vary the amplitude of the initial power spectrum. Specifically, for each $\delta_{L0}$ value for which we simulate a separate universe, we also simulate the fiducial cosmology with the initial power spectrum amplitude multiplied by $\tilde D(t_0)^2 /D(t_0)^2$, where $\tilde D(t_0)$ is the linear growth factor in the corresponding separate universe cosmology. The results of these simulations are shown in \refsec{rescaled-amplitude}.

%%%%%%%%%%%%%%%%%%%%%%%%%%%%%%%%%%%%%%%%%%%%%%%%%%%%%%%%%%%%%%%%%%%%%%%%%%%
%%%%%%%%%%%%%%%%%%%%%%%%%%%%%%%%%%%%%%%%%%%%%%%%%%%%%%%%%%%%%%%%%%%%%%%%%%%
\section{Results}
\label{sec:results}

For the power spectrum computation, we first estimate the density
contrast $\delta({\bf x})$ on a $1024^3$ grid using the cloud-in-cell
mass assignment scheme, then apply a Fast Fourier transform, and angular
average the squared amplitude $|\delta_{\bf k}|^2$. The density contrast
$\delta({\bf x})=\rho({\bf x})/\bar \rho-1$ describes the overdensity
with respect to the reference density $\bar\rho$. When we are interested
in the growth-only response function, $\bar\rho$ is equal to the mean
density of the separate universe. When we compute the full response
function, $\bar\rho$ is equal to the mean density of the fiducial cosmology.
Similarly, for the growth-only response, distances are measured using the comoving coordinates of the respective cosmology\footnote{Note, however, that the unit of length is always $h^{-1}$Mpc, where $h$ corresponds to the fiducial cosmology.}, whereas, for the full response, the power spectrum is always measured in comoving coordinates of the fiducial cosmology. 

We only report results up to a maximum wavenumber of $2\,h^{-1}$Mpc. A convergence study with simulations with 8 times lower mass resolution shows differences in $G_1$, $G_2$ and $G_3$ of only 1 (3) to 5 (10) percent at $z=0$ ($z=2$) up to that wavenumber, where the deviations increase from the linear response function to the higher-order response functions. The results for the full response functions $R_1$, $R_2$ and $R_3$ are converged to an even better degree.
We therefore expect that the simulation results presented in this paper are converged to a sub-percent to a few percent level.

In order to compute the first three response functions, we fit 
a polynomial in $\delta_L$ to the fractional difference in the measured power spectrum $\Delta_k(\delta_L)\equiv P(k|\delta_L)/P(k|\delta_L=0)-1$ 
for each $k$-bin.  For the fit, we only include results from separate universe simulations with $|\delta_L(t_{\rm out})| \le 0.5$ and use a polynomial with degree 6 to be unbiased from higher-order response functions.  
As the random realization of the initial density field is the same across different $\delta_L$ values, the corresponding power spectra are strongly correlated. By considering the ratio, or the relative difference, of two power spectra a large fraction of the noise cancels. 
However, for the same realization, the measured fractional differences $\Delta_k(\delta_L)$ are still correlated over $\delta_L$. 
As the number of realizations (16) is not large enough to reliably estimate the covariance between different $\delta_L$ values, we cannot include this correlation in the polynomial fitting. Instead, we construct quasi-decorrelated samples of $\Delta_k(\delta_L)$ by randomly choosing a realization for each $\delta_L$ value. 
Fitting many of those subsamples allows for a robust error estimation of the derived response functions.

%%%%%%%%%%%%%%%%%%%%%%%%%%%%%%%%%%%%%%%%%%%%%%%%%%%%%%%%%%%%%%%%%%%%%%%%%%%
\subsection{Growth-only response functions}
\label{sec:growth-only}

\begin{figure}[t]
\includegraphics[clip=false,trim= 0cm 0cm 0cm 0.cm, angle=-90,width=0.48\textwidth]{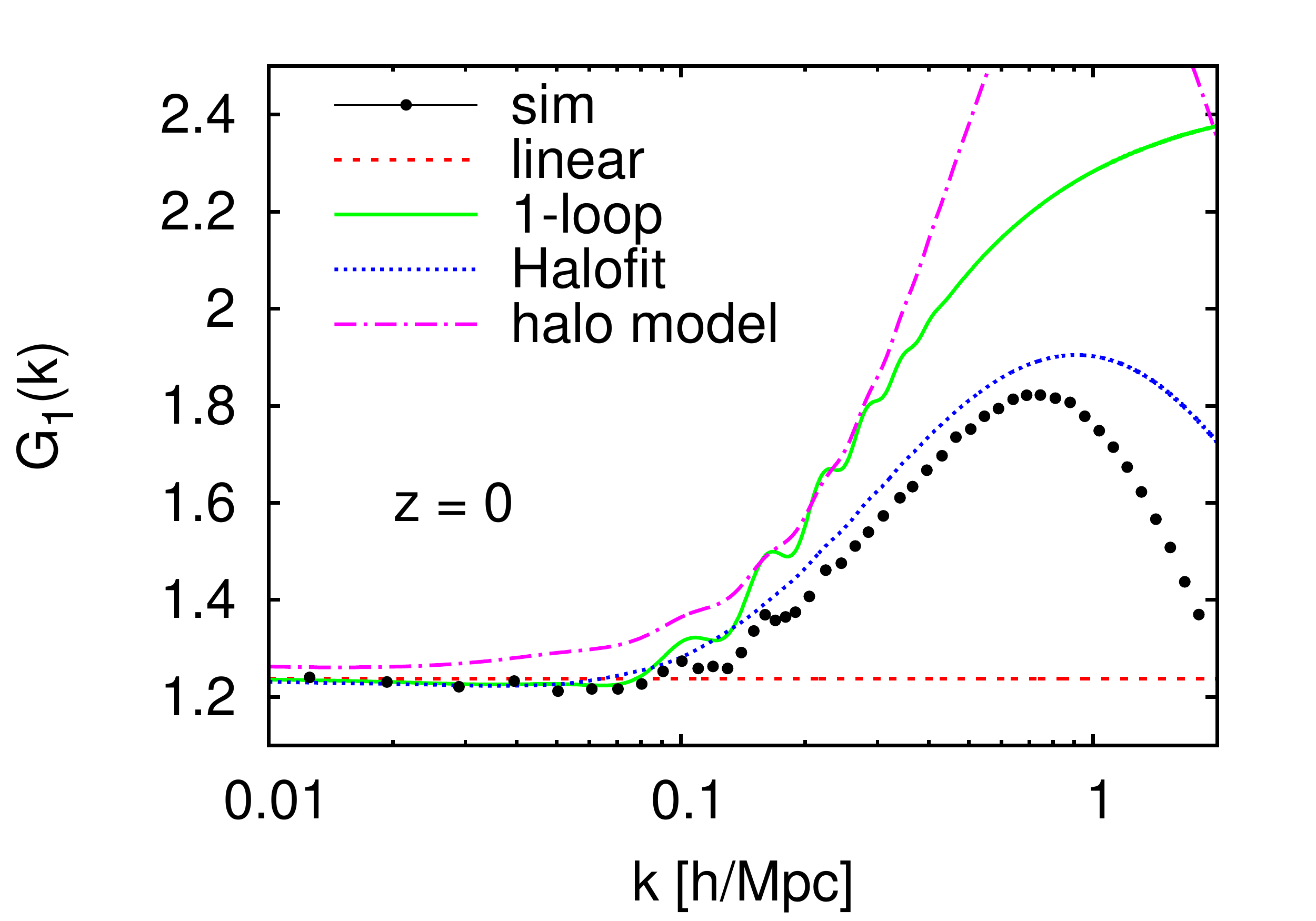}
\includegraphics[clip=false,trim= 0.cm 0.cm 0.cm 0.cm, angle=-90,width=0.48\textwidth]{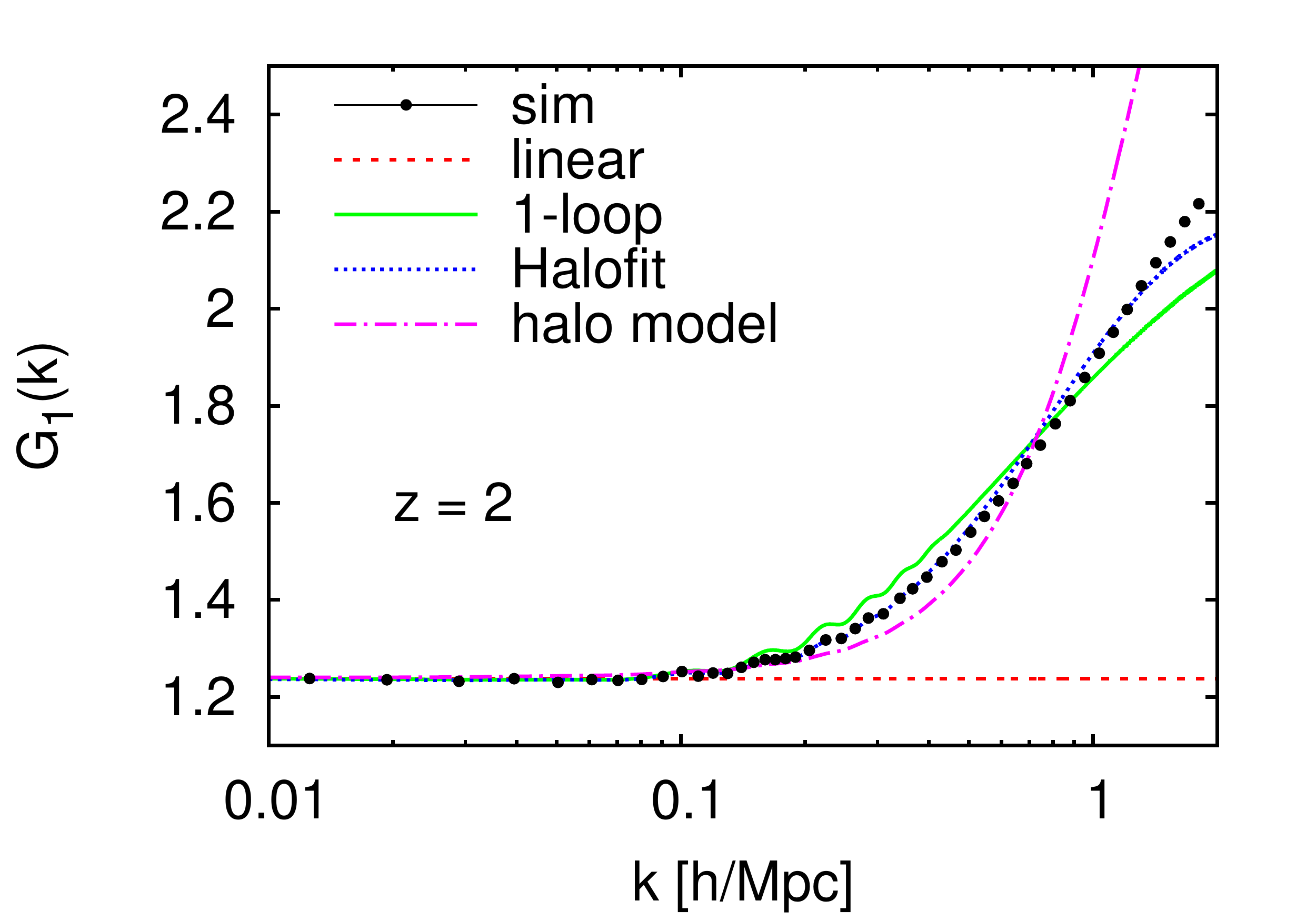}
\includegraphics[clip=false,trim= 0cm 0cm 0cm 0.cm,angle=-90,width=0.48\textwidth]{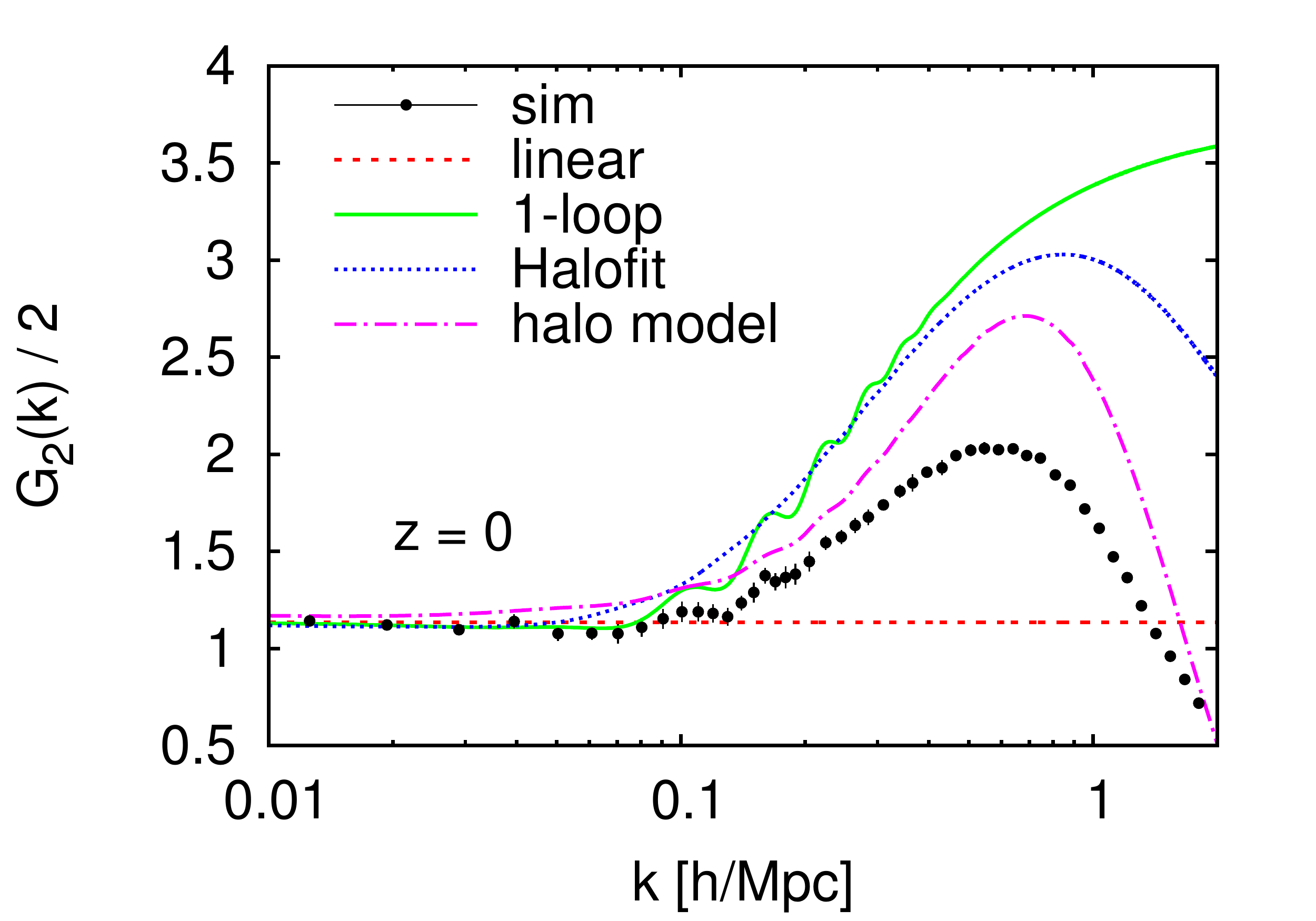}
\includegraphics[clip=false,trim= 0.cm 0.cm 0.cm 0.cm,angle=-90,width=0.48\textwidth]{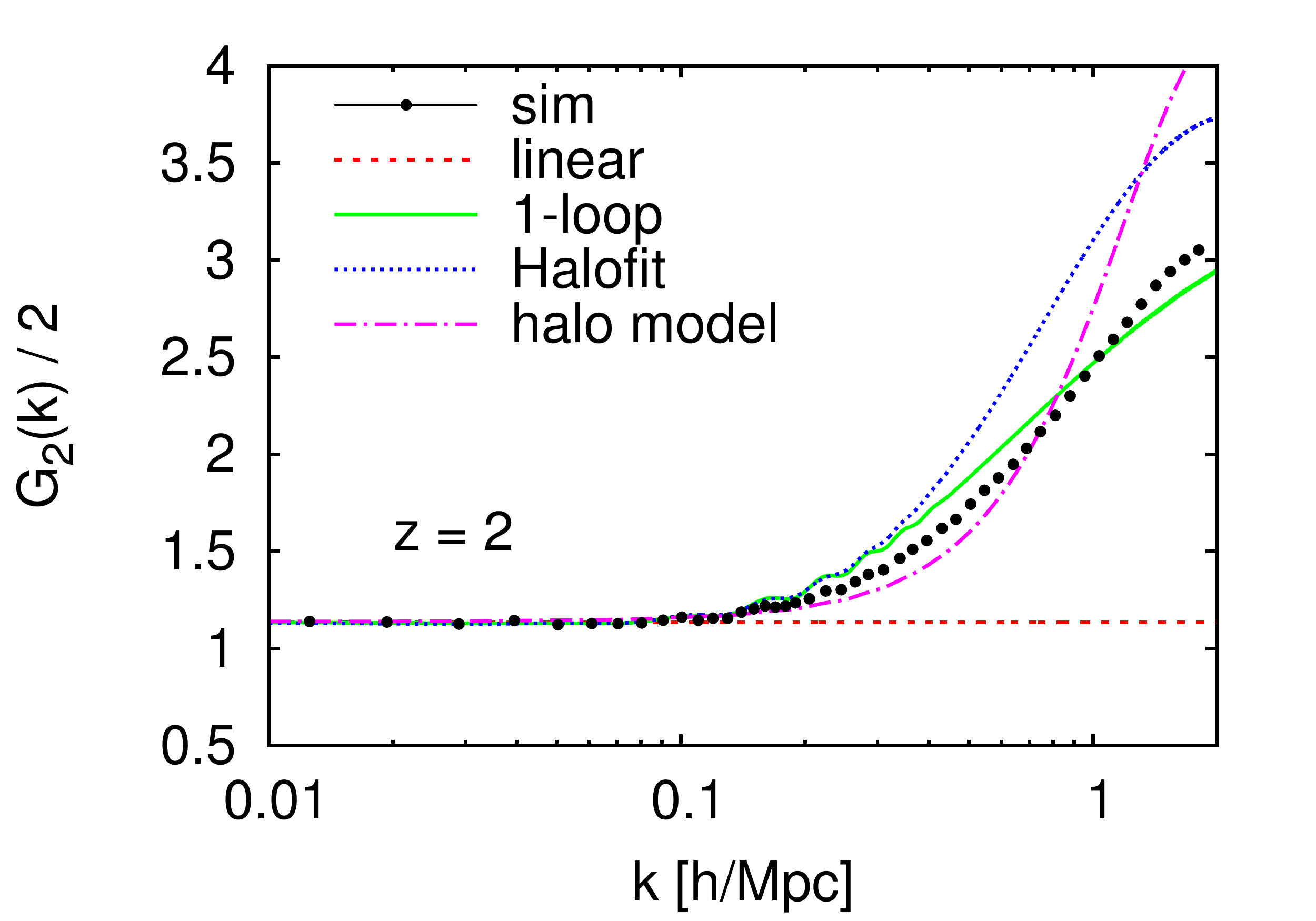}
\includegraphics[clip=false,trim= 0cm 0cm 0cm 0cm,angle=-90,width=0.48\textwidth]{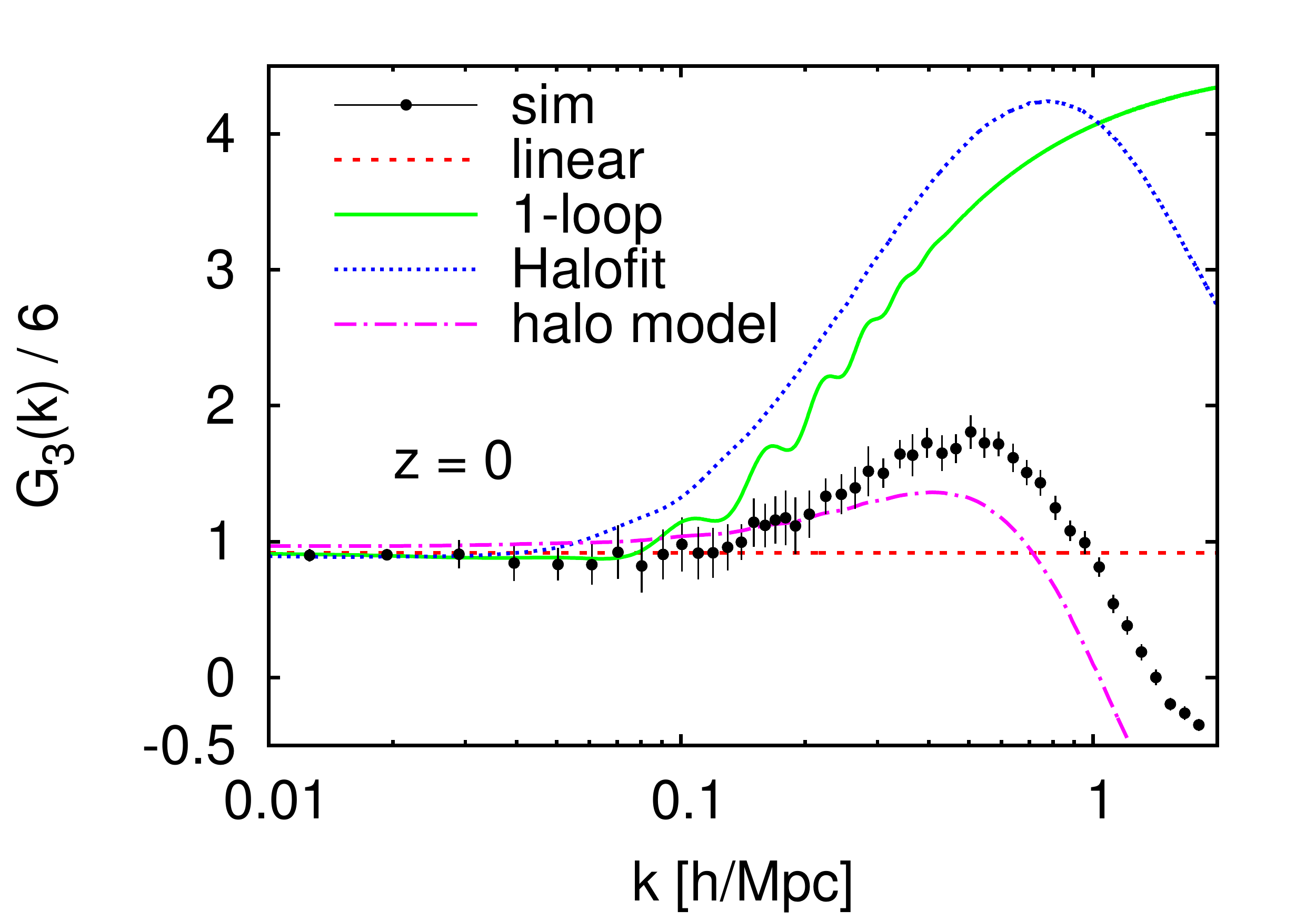}
\hspace{0.4cm}
\includegraphics[clip=false,trim= 0.cm 0.cm 0.cm 0.cm,angle=-90,width=0.48\textwidth]{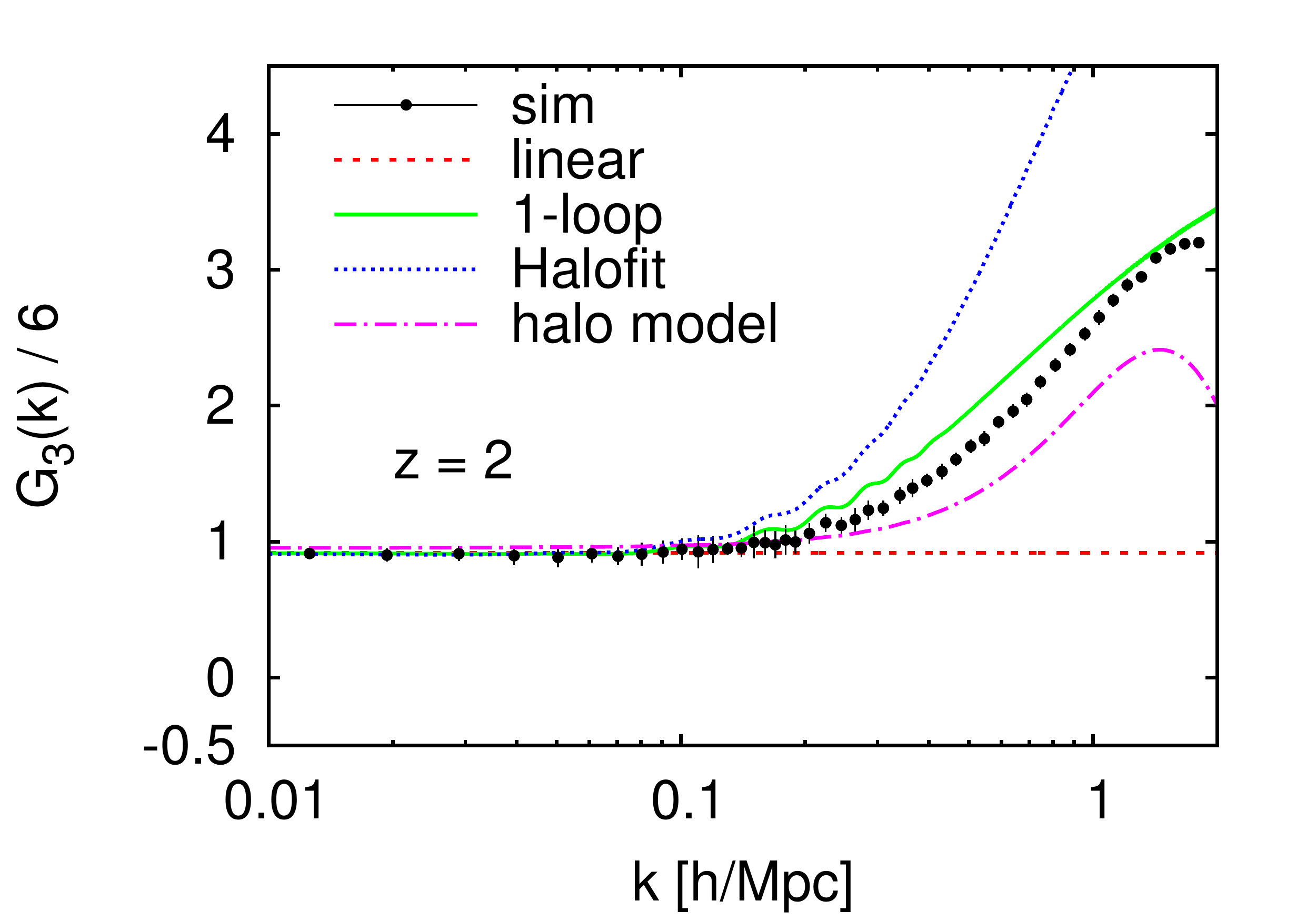}
 \caption{The first three growth-only response functions of the power spectrum measured from the separate universe simulations at $z=0$ (left) and $z=2$ (right). 
 The error bars show the statistical error derived by random
resampling of the data (see text). For data points apparently without error bars, the statistical error is
 smaller than the size of a dot.}
  \label{fig:growth_only}
\end{figure}

\refFig{growth_only} shows the first three growth-only response
functions measured from the simulations at $z=0$ (left column) and $z=2$
(right column). These correspond to the fully nonlinear squeezed limit
bispectrum (3-point function), trispectrum (4-point function) and
5-point function, and are essentially the same as in
figure~2 of Ref.~\cite{sep1}, although the results presented there 
are derived from lower resolution simulations (particle load of
$256^3$). 
The small wiggles in the growth-only response functions result from the damping of the baryon acoustic oscillations (BAO), which depends on the amplitude of density fluctuations and thus on $\delta_L$. 

Let us compare the simulation results to the theoretical predictions discussed in \refsec{response}.  
On sufficiently large scales, the perturbation theory predictions are the most accurate, as expected.  At high redshift, the 1-loop predictions best describe the results overall.  The 1-loop predictions also show a BAO damping effect.  
At $z=0$, the growth-only response is captured best by the \textsc{halofit} prescription
(in case of $G_1$) or the halo model (in case of $G_2,\,G_3$).  
We see that the \textsc{halofit} prescription describes the simulation results of the linear response well at both redshifts, but performs significantly worse for the higher-order response functions.  The BAO damping effect is essentially absent in both \textsc{halofit} and halo model predictions.  
Overall, none of the models is able to accurately describe the simulation data in the nonlinear regime, with discrepancies at $z=0$ ranging from 20\% in the best case to a factor of several.  
These discrepancies are not surprising given that we are looking at scales
beyond the validity of perturbation theory and at higher $N$-point functions
for which the semi-analytical approaches were not tuned.  

The halo model prediction does not asymptote exactly to the linear
result in the $k\to 0$ limit.  
This is because the 1-halo term asymptotes
to a white noise contribution in this limit, and since the 1-halo term
contributes to $G_n$ due to the dependence of the halo mass function on $\delta_L$ 
(\refsec{halomodel}), this induces a correction to the linear prediction which
contributes on large scales.  Physically, this occurs because the halo model does
not enforce momentum conservation of the matter density field. This issue
can be fixed by introducing a ``mass compensation scale'' \cite{mohammed/seljak}.  

The halo model predictions can be tuned to better match the simulation
results by allowing for a dependence of the halo profiles on the long-wavelength density, which is expected on physical grounds (see also \cite{li/hu/takada/3}).  Specifically, if the
the scale radius of halos at fixed mass increases in the presence of a long-wavelength
density perturbation, this lowers the peak in the response and thus could
lead to better agreement with the simulations results.  A detailed investigation of this is beyond the scope of the present paper.

%%%%%%%%%%%%%%%%%%%%%%%%%%%%%%%%%%%%%%%%%%%%%%%%%%%%%%%%%%%%%%%%%%%%%%%%%%%
\subsection{Comparison to simulations with rescaled initial amplitude}
\label{sec:rescaled-amplitude}

All models for the growth-only response functions that we have presented
in \refsec{response} and shown above are based on the approximation that
we can trade the effect of $\delta_L$ for an appropriate change to the linear growth factor (or equivalently, the linear power spectrum).   But how well does this approximation work?   
Using the set of simulations described in \refsec{growth_history}, we
can explicitly test this approximation on all scales including the
nonlinear regime.  

\begin{figure}[t]
\includegraphics[clip=false,trim= 0cm 0cm 0cm 0.cm, angle=0,width=0.48\textwidth]{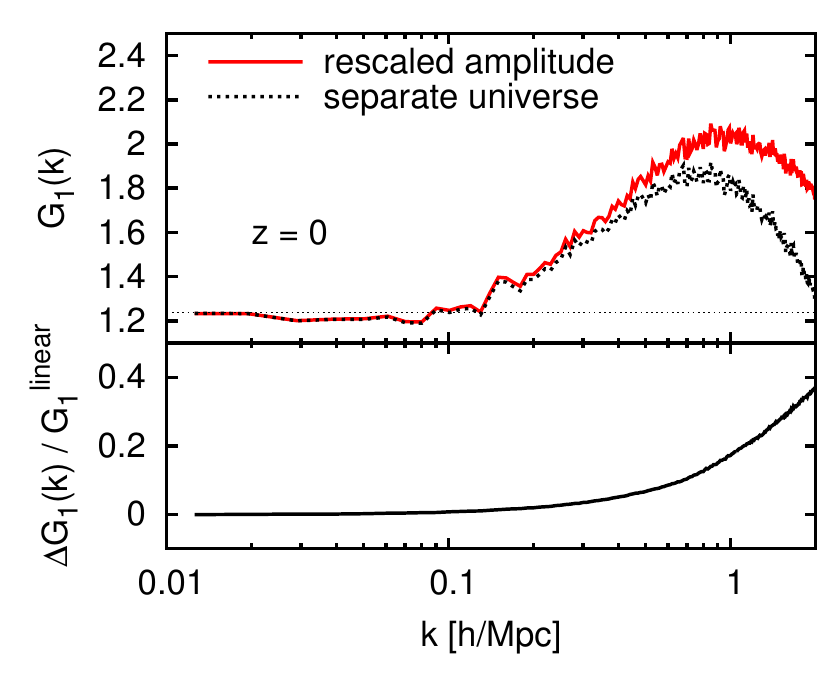}
\includegraphics[clip=false,trim= 0cm 0cm 0cm 0.cm, angle=0,width=0.48\textwidth]{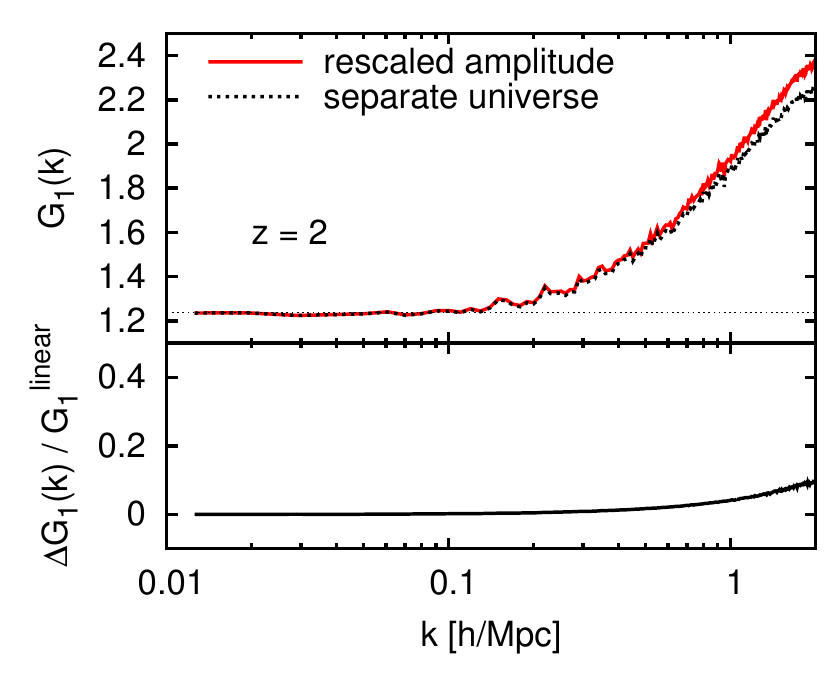}
\includegraphics[clip=false,trim= 0cm 0cm 0cm 0.cm,angle=0,width=0.48\textwidth]{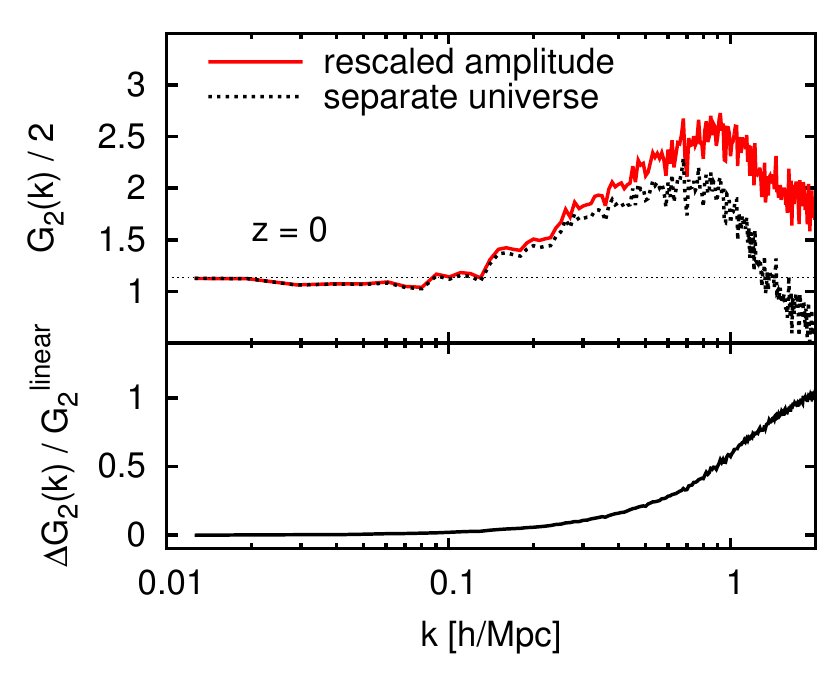}
\includegraphics[clip=false,trim= 0cm 0cm 0cm 0.cm,angle=0,width=0.48\textwidth]{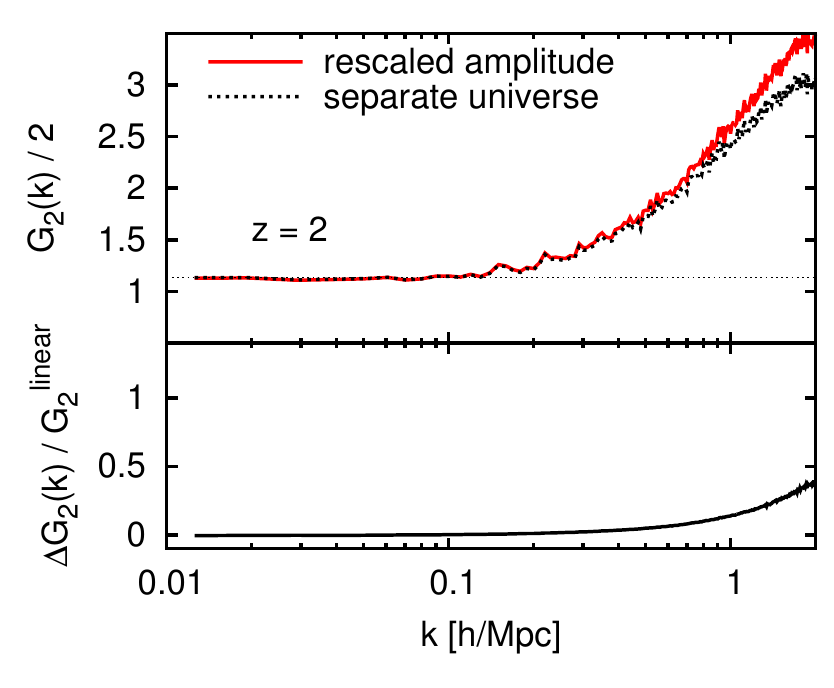}
\includegraphics[clip=false,trim= 0cm 0cm 0cm 0.cm,angle=0,width=0.48\textwidth]{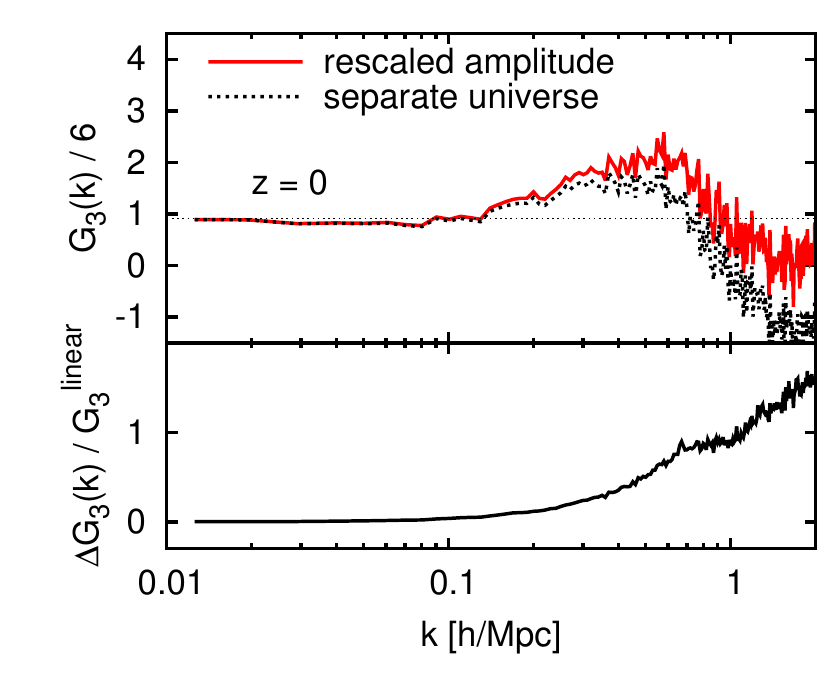}
\hspace{0.4cm}
\includegraphics[clip=false,trim= 0cm 0cm 0cm 0.cm,angle=0,width=0.48\textwidth]{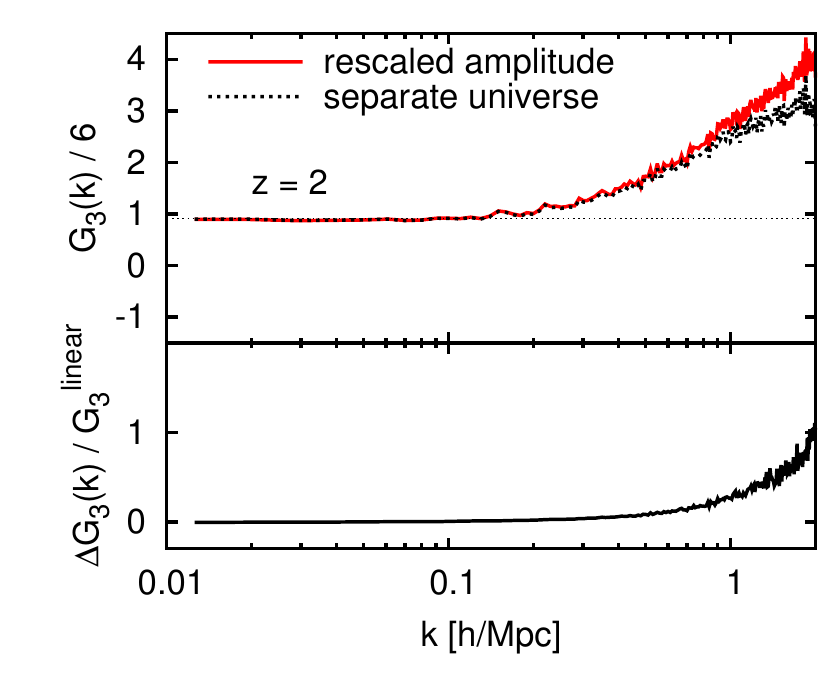}
 \caption{Comparison of the growth-only response functions $G_1,\,G_2,\,G_3$ (top to bottom)
 measured at $z=0$ (left column) and $z=2$ (right column) from
 one realization of the separate universe simulations and from the same
 realization simulated by varying the initial amplitude. The bottom
 sub-panels show the difference, $\Delta G_n=G_n^{\rm rescaled}-G_n^{\rm
 separate}$, divided by the response of the linear matter power
 spectrum, $G_n^{\rm linear}$.}
  \label{fig:growth_comparison}
\end{figure}

In \reffig{growth_comparison}, we show the growth-only response
functions measured from two different sets of simulations.  In case of
$G_1$, this comparison was also shown in figure~6 of \cite{li/hu/takada/3},
and our results agree with theirs.\footnote{Note that the authors of ref.~\cite{nishimichi/valageas:2014} perform a different comparison using the time derivative of the nonlinear power spectrum in simulations of the fiducial cosmology.} The
``rescaled amplitude simulations'' of \refsec{growth_history} all assume
the fiducial cosmology but vary the amplitude of the linear power
spectrum used to initialize the simulations so as to match the
linear power spectrun in the modified cosmology at the given output times 
[\refeq{Dtilde} and \refeq{growth_approx}].  
On linear scales, these simulations thus agree with the ``separate universe''
simulations by construction.   
As the simulations share the same random realization of the initial
density, the sample variance (noise in the upper panels) gets vastly
reduced when considering the difference of the measured response
functions, $\Delta G_n=G_n^{\rm rescaled}-G_n^{\rm separate}$.  This is
shown in the lower subpanels of \reffig{growth_comparison}, where we
have divided $\Delta G_n$ by the corresponding linear growth-only response, i.e. the prediction in the $k\to 0$ limit.  

The differences seen in \reffig{growth_comparison} are caused by the different growth history, which is not captured by the rescaling of the initial amplitude.   
Following the discussion in \refsec{NLgrowth}, the commonly used
SPT approach factorizing the growth factor and scale dependence assumes at 
all orders that
a long wavelength density perturbation enters exclusively through the 
modified linear growth.  Thus, {\it even when calculated to all orders}, the best
that this SPT calculation could do is to reproduce the rescaled amplitude result in \reffig{growth_comparison}, which deviates from the actual response at $z=0$ by 10\% at $k \simeq 0.5 \iMpch$ and 20\% at $k\simeq 1 \iMpch$ for $G_1$, and significantly worse
for the higher-order response functions.  At $z=2$ on the other hand, the rescaled-amplitude $G_1$ matches the separate universe response to better than 10\% even beyond $k = 1 \iMpch$, and for $G_2,\,G_3$ performs significantly better as well.  

There are two possible explanations for these discrepancies in the SPT context.  
First, using the SPT kernels derived for an Einstein-de Sitter
universe (which have time-independent coefficients), with the $\Lambda$CDM linear growth factor replacing the Einstein-de Sitter $a(t)$, could become highly inaccurate for $\Lambda$CDM at higher orders.  Note that the same issue exists for a fiducial flat Einstein-de Sitter universe, since for $\delta_L\neq 0$ the quantity $\Omega_m/f^2$ is no longer 1 (in fact, $d(\Om/f^2)/d\delta_L=-5/21$ \cite{ben-dayan}; see also the discussion in \cite{nishimichi/valageas:2014}).  There is no indication of such a strong effect at low orders in perturbation theory, where this approximation typically performs to better than a percent \cite{bernardeau/etal:2001}.   Furthermore, ref.~\cite{ben-dayan} found that a cancelation in the curvature contribution to the growth integral suppresses this effect.  Finally, ref.~\cite{li/hu/takada/3} shows that the growth-only response of the power spectrum to a change in the Hubble constant while keeping $\Omega_m h^2$ fixed follows the separate universe response very closely (Fig.~6 there).  If the much larger discrepancies between separate universe response and rescaled amplitude response were due to the cosmology dependence of the SPT kernels, one would not expect this to be the case.  Nevertheless, we do not claim to be able to rigorously exclude this possibility.  

The other possibility, more likely in our opinion, is that the discrepancy
between rescaled amplitude and full separate universe simulations is due to
effective non-perfect fluid terms, such as pressure and anisotropic stress, in the 
dark matter fluid \cite{baumann/etal}.  The effective fluid properties 
depend on highly nonlinear small scales which are not described by the 
Euler-Poisson system.  Their value can depend on the growth history (as well as
the power spectrum shape) thus leading to a discrepancy between
rescaled amplitude and separate universe simulations.  Assuming this
interpretation is correct, \reffig{growth_comparison} explicitly
shows the breakdown of SPT on nonlinear scales as effective pressure,
anisotropic stress and sound speed need to be included.  Separate universe
simulations can be used to measure the response of these effective terms to
a long-wavelength overdensity, which is crucial when modeling $(N>2)$-point functions.
The results shown in \reffig{growth_comparison} are analogous to what 
has been found for the mass function of halos which is a key ingredient in the halo model description of the nonlinear matter density field.  
The mass function shows departures from being a simple function of the linear matter power spectrum at the 5--10\% level \cite{tinker/etal,mf}.  

In an Einstein-de Sitter cosmology with scale-invariant initial power
spectrum $P_l(k) \propto k^n$, there is only one characteristic spatial scale at any given time,
which corresponds to the scale at which the density field becomes order
1 \cite{pajer/zaldarriaga}.  Let us denote this wavenumber as $k_{\rm NL}(t)$.  Then,
the response functions have to follow a universal function
of $k/k_{\rm NL}(t)$, i.e. $G_m(k, t) = G_m(k/k_{\rm NL})$ (keeping the index
of the initial power spectrum fixed).  Thus, in this specific 
case, separate universe simulations and rescaled-amplitude simulations
will give exactly the same result when compared at fixed $k/k_{\rm NL}$.  
The departures shown in \reffig{growth_comparison} can thus be seen as a consequence
of the $\Lambda$CDM background and the departure from scale-invariance of
the initial power spectrum.  It would be interesting to disentangle the
two effects, e.g. by performing separate universe simulations in $\Lambda$CDM
with scale-invariant initial conditions.  We leave this for future work,
but point out that when plotting the differences shown in the lower panel
of \reffig{growth_comparison} as a function of $k/k_{\rm NL}$, we still find a factor of
several difference in the $z=0$ and $z=2$ results.

Note that in Ref.~\cite{mohammed/seljak}, the authors performed a similar comparison as the one shown for $G_1$ in \reffig{growth_comparison}.  
However, instead of studying the growth-only response as considered here,
they included the reference density effect and found a significantly larger
discrepancy between the two measurements considered.  
As discussed in \refsec{response}, however, 
the reference density effect is a simple remapping, and the growth-only
response considered here is the proper quantity to compare for the question
we are interested in, namely to measure the effect of the growth history.  

\begin{figure}
\includegraphics[clip=false,trim= 0cm 0cm 0cm 0.cm, angle=-90,width=0.48\textwidth]{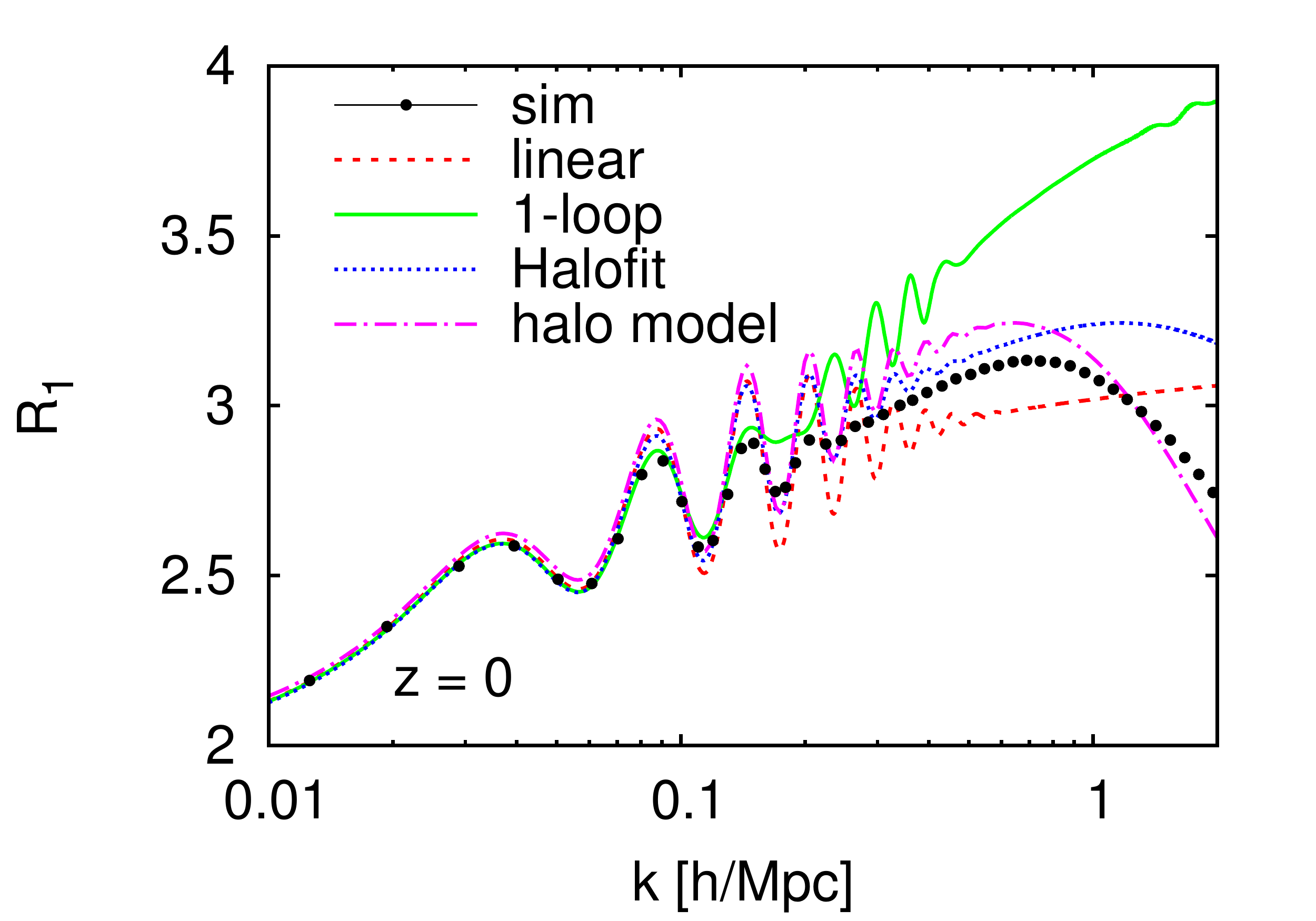}
\includegraphics[clip=false,trim= 0.cm 0.cm 0.cm 0.cm,angle=-90,width=0.48\textwidth]{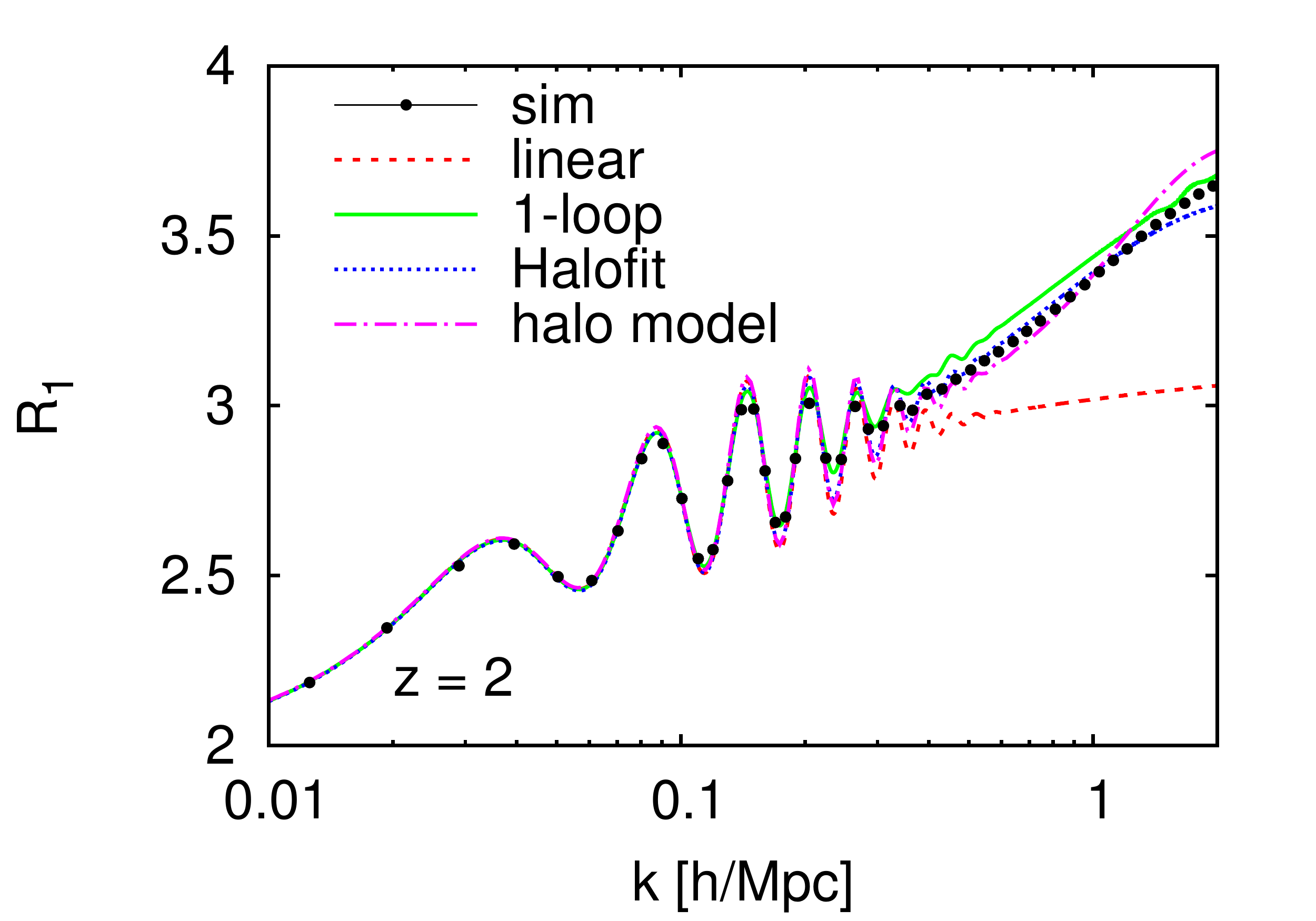}
\includegraphics[clip=false,trim= 0cm 0cm 0cm 0.cm,angle=-90,width=0.48\textwidth]{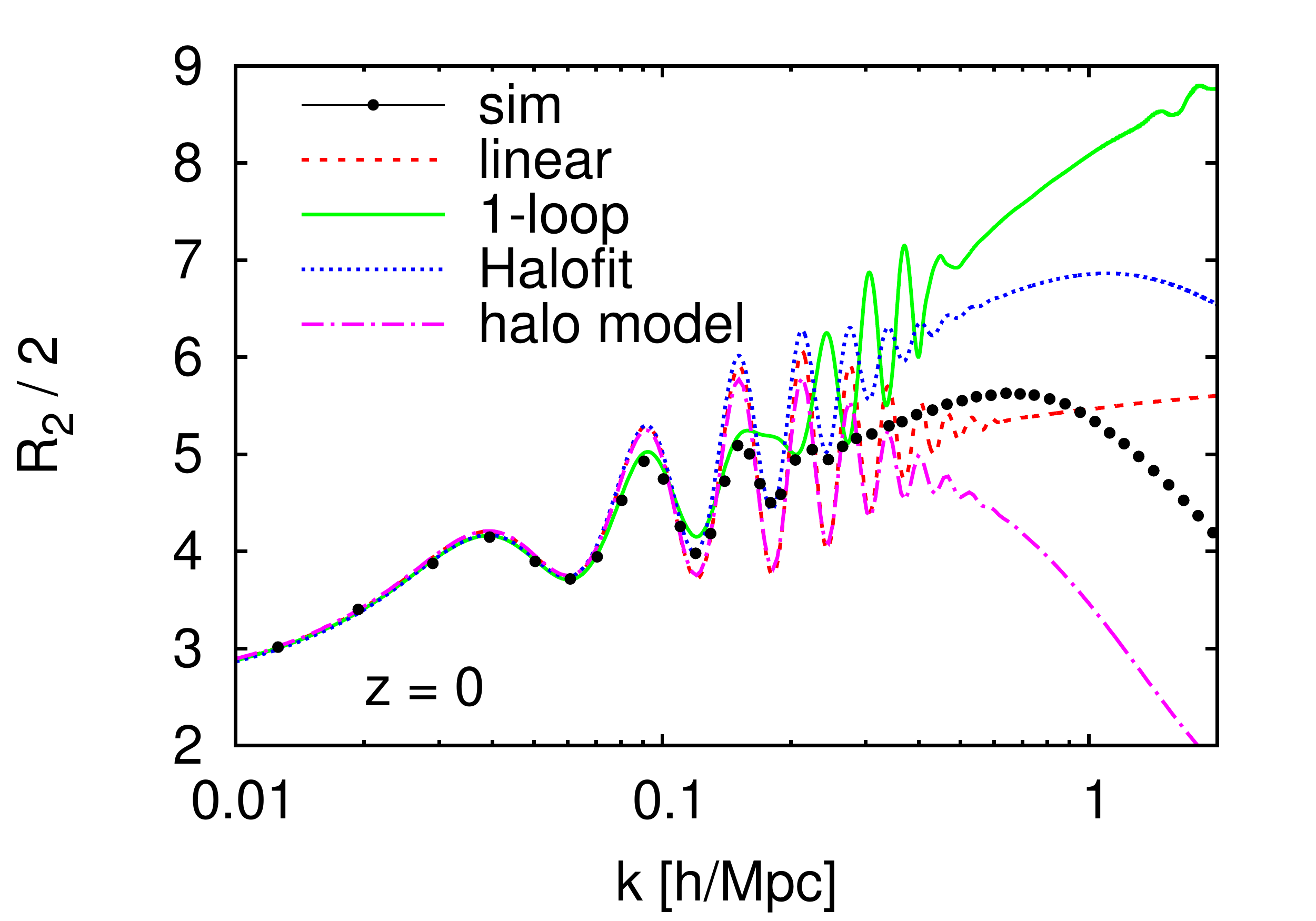}
\includegraphics[clip=false,trim= 0.cm 0.cm 0.cm 0.cm,angle=-90,width=0.48\textwidth]{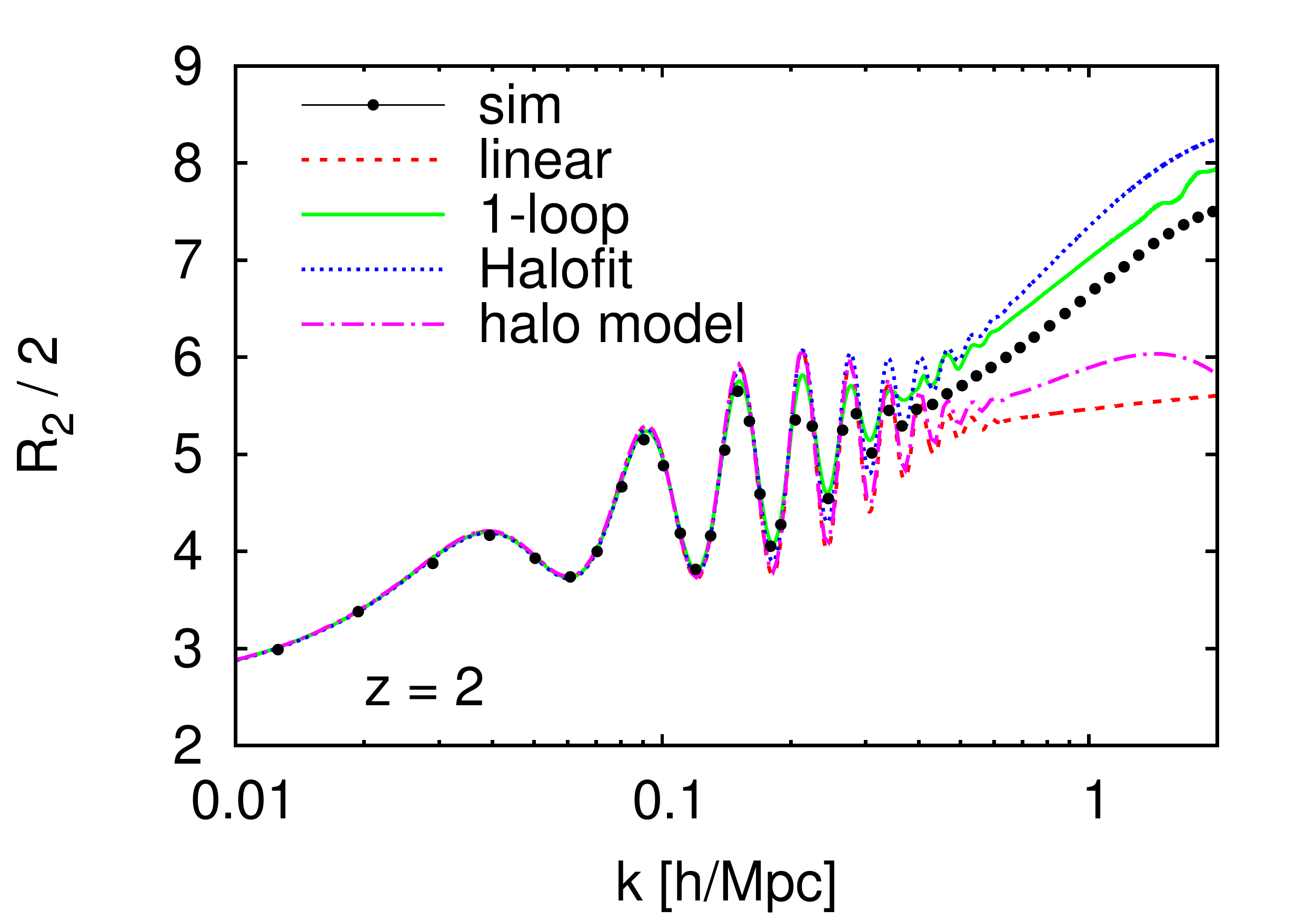}
\includegraphics[clip=false,trim= 0cm 0cm 0cm 0cm,angle=-90,width=0.48\textwidth]{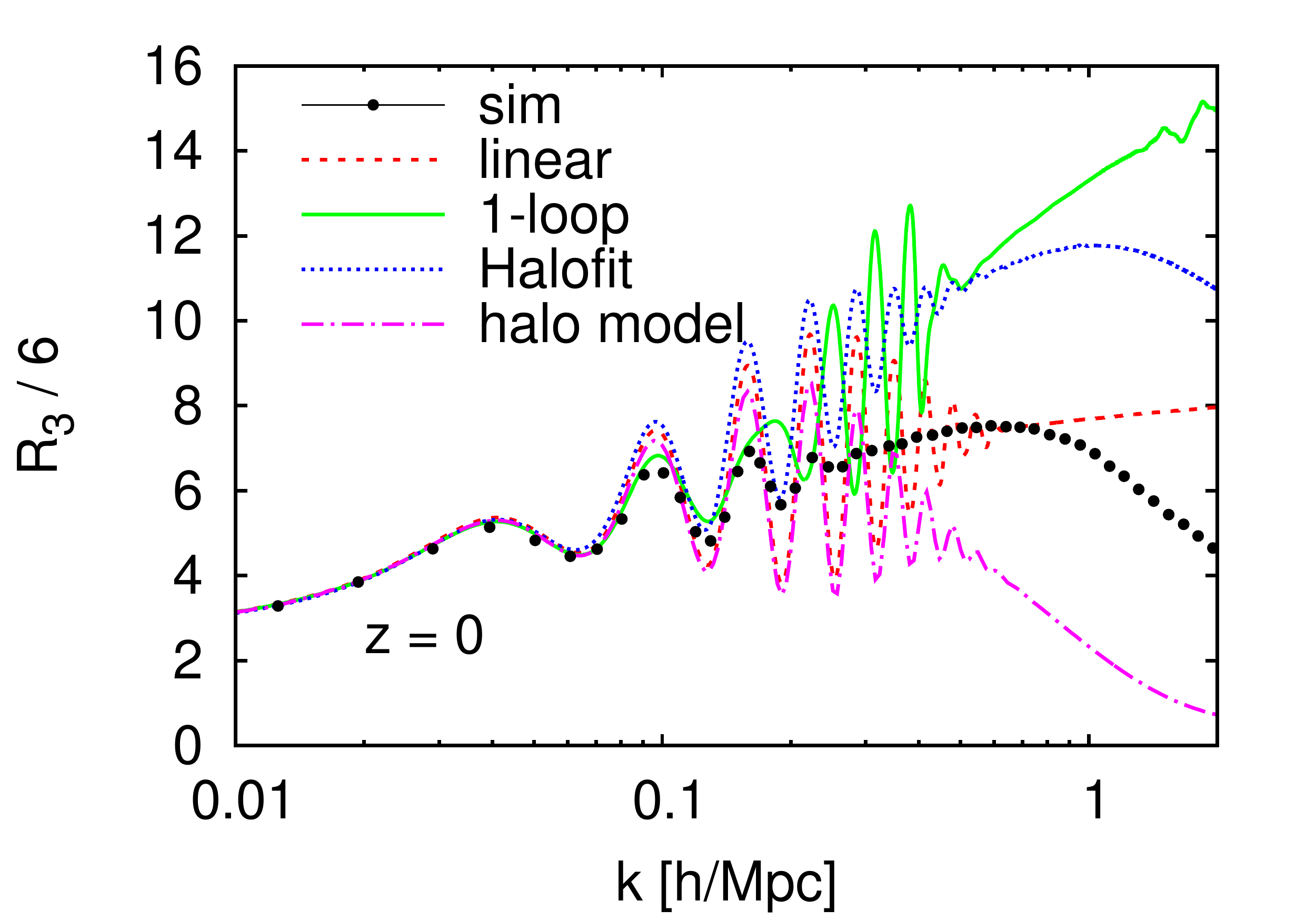}
\hspace{0.4cm}
\includegraphics[clip=false,trim= 0.cm 0.cm 0.cm 0.cm,angle=-90,width=0.48\textwidth]{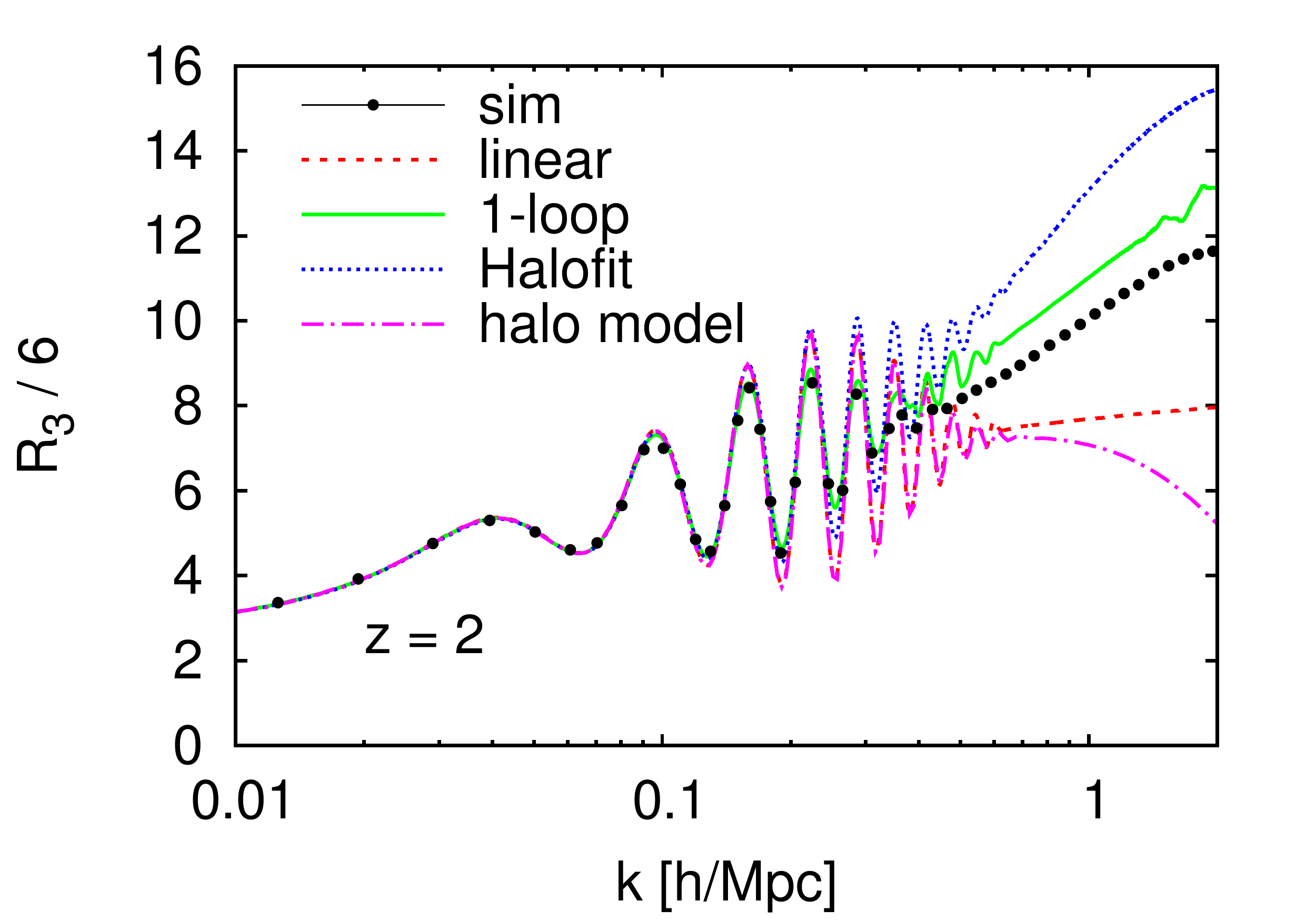}
 \caption{The first three full response functions of the power spectrum measured from the separate universe simulations at $z=0$ (left) and $z=2$ (right). }
  \label{fig:full}
\end{figure}

%%%%%%%%%%%%%%%%%%%%%%%%%%%%%%%%%%%%%%%%%%%%%%%%%%%%%%%%%%%%%%%%%%%%%%%%%%%
\subsection{Full response functions}
\label{sec:full}

We now turn to the results for the full response functions, i.e.~including the ``dilation'' and ``reference density'' effects.
The results of the simulations and the model predictions are shown in \reffig{full}.  The oscillations in the response functions can be traced back to the BAOs in the power spectrum.  The BAOs propagate to the response functions primarily by the ``dilation'' effect, which yields derivatives of the power spectrum with respect to $k$ (see \refeqs{R1}{R3}).  
The 1-loop perturbation theory predictions describe the simulation results accurately up to $k\le 0.15\,h^{-1} {\rm Mpc}$ and $k\le 0.3\,h^{-1} {\rm Mpc}$ at $z=0$ and $z=2$, respectively. 
As the other theoretical models do not include the damping of the BAOs in the nonlinear power spectrum, they predict oscillations in the response functions which are too large. To improve the accuracy of those models around the BAO scale, one would need to put in the BAO damping by hand. In the nonlinear regime, none of the models is able to reproduce the simulation data.
In principle, one could build a hybrid model for the full response by combining an accurate prediction of the nonlinear power spectrum of the fiducial cosmology and the growth-only response functions $G_n(k)$ discussed in the previous section. In this paper, however, we do not pursue this approach.

\begin{figure}
\includegraphics[clip=false,trim= 0cm 0cm 0cm 0.cm, angle=-90,width=0.48\textwidth]{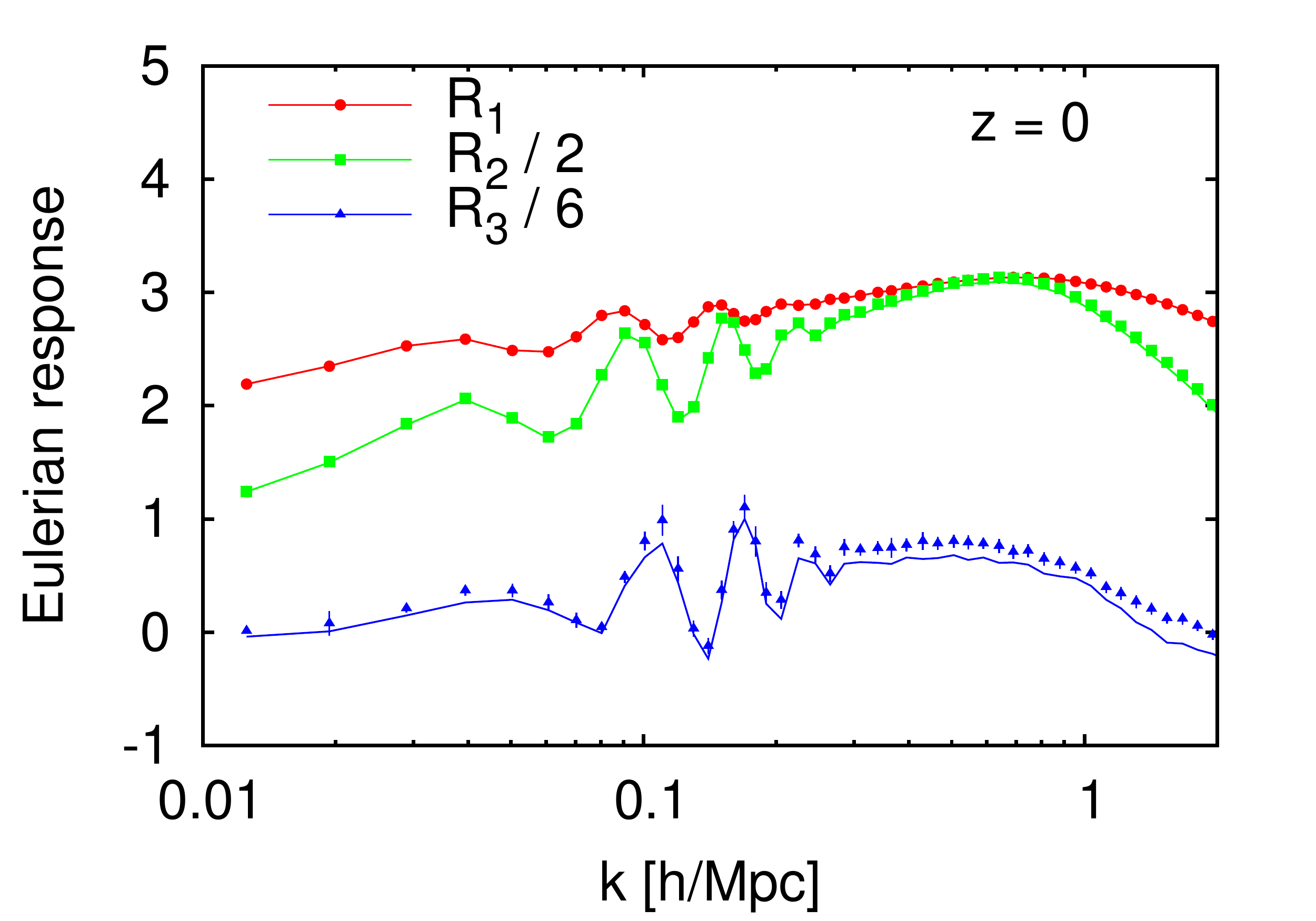}
\includegraphics[clip=false,trim= 0.cm 0.cm 0.cm 0.cm, angle=-90,width=0.48\textwidth]{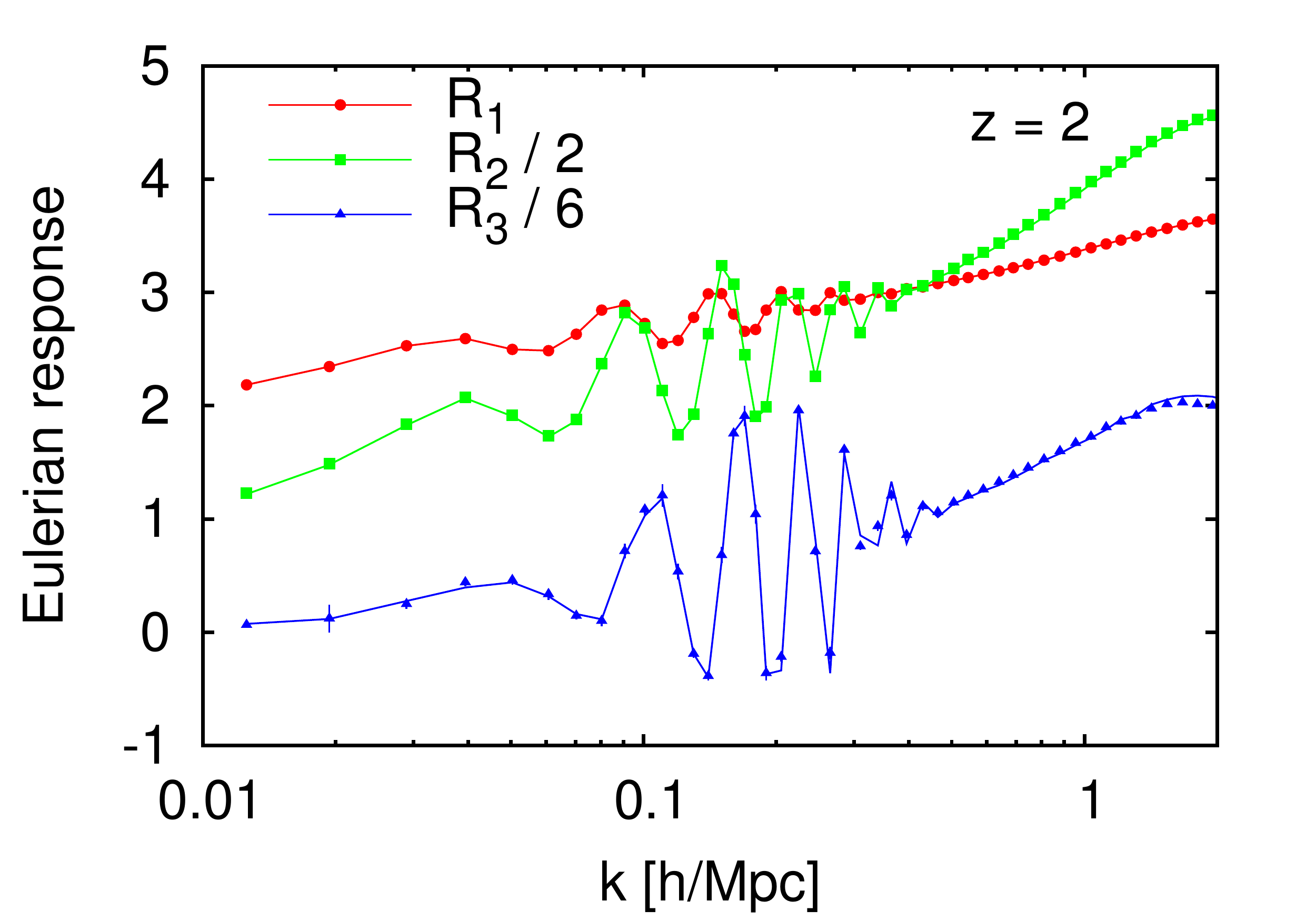}
 \caption{The first three Eulerian response functions of the power spectrum measured from the separate universe simulations (data points) at $z=0$ (left) and $z=2$ (right). The lines show the corresponding linear combinations of the Lagrangian response functions using the $f_n$ coefficients derived for the Einstein-de\ Sitter universe (see \refeq{Reuler} and \refeq{fn}).}
  \label{fig:eulerian_response}
\end{figure}

%%%%%%%%%%%%%%%%%%%%%%%%%%%%%%%%%%%%%%%%%%%%%%%%%%%%%%%%%%%%%%%%%%%%%%%%%%%
\subsection{Eulerian response functions}

So far, we have always considered the response to the linearly-extrapolated initial
(Lagrangian) overdensity $\delta_L$.   We now consider the corresponding response to the \emph{evolved nonlinear}
(Eulerian) overdensity $\delta_\rho$. Using the expansion derived for the Einstein-de Sitter universe, \refeq{fn}, we find
\ba
R_1^{\rm Eulerian}(k)&= R_1(k)\,, \vs
R_2^{\rm Eulerian}(k)&= R_2(k) - 2 f_2R_1(k)\,, \vs
R_3^{\rm Eulerian}(k)&= R_3(k) -6 f_2 R_2(k)+6\left(2 f_2^2-f_3\right)R_1(k)\,.
\label{eq:Reuler}
\ea
In \reffig{eulerian_response}, we compare the directly measured Eulerian response functions with the appropriate linear combinations of the measured Lagrangian response functions. The agreement is excellent as expected, especially at high redshift at which the $\Lambda$CDM universe is very well approximated by the Einstein-de Sitter universe.  

Interestingly, the higher-order Eulerian response functions are much smaller than in the Lagrangian case.  That is, the response of the nonlinear matter
power spectrum to a uniform nonlinear final-time density $\delta_\rho$ is
close to linear.  This is most likely due to the fact that the growth-only response functions are subdominant compared to the rescaling and reference density contributions, especially at higher order.  In this case, \refeq{refdensity_dilation} implies a close to linear scaling with $\delta_\rho$.

%%%%%%%%%%%%%%%%%%%%%%%%%%%%%%%%%%%%%%%%%%%%%%%%%%%%%%%%%%%%%%%%%%%%%%%%%%%%%
%%%%%%%%%%%%%%%%%%%%%%%%%%%%%%%%%%%%%%%%%%%%%%%%%%%%%%%%%%%%%%%%%%%%%%%%%%%%%
\section{Conclusions}
\label{sec:conclusion}
In this paper, we employed dedicated N-body simulations using the separate universe technique presented in
Ref.~\cite{sep1}  to compute the response of the nonlinear matter power spectrum
to a homogeneous overdensity superimposed on a flat FLRW universe. The
response functions we computed give the squeezed limits of the 3-, 4-, and
5-point functions, in which all but two wavenumbers are taken to be small
and are angle-averaged.  
By virtue of the separate universe technique, we reach an unprecedented 
accuracy of these nonlinear matter $N$-point functions.

The response function consists of three parts: changing the reference
density with respect to which the power spectrum is defined; rescaling
of comoving coordinates; and the effect on the growth of structure. The
former two effects can be calculated trivially, whereas the third one
requires separate universe simulations. We have compared the simulation
results with analytical and semi-analytical results, in particular
standard perturbation theory (SPT),
the empirical fitting function \textsc{halofit}, and the halo model, finding that SPT 
typically yields the best results at high redshifts.  The fitting 
function and halo model, while qualitatively describe the trends
seen in the response functions, give a poor quantitative description 
on nonlinear scales.  

A fundamental assumption of all of the analytical and semi-analytical methods
used in this paper, including standard perturbation theory at any order, is
that nonlinear matter statistics at a given time are given solely by the
linear power spectrum at the same time, and do not depend on the growth
history otherwise. As was done in \cite{li/hu/takada/3} for the response
function for $n=1$, we were able to test this assumption for $n=2$ and 3
quantitatively by comparing the separate universe simulations with
simulations with a rescaled initial power spectrum amplitude.   We find
that this assumption fails at the level of 10\% at $k\simeq 0.2 -
0.5 \iMpch$ for $5$- to $3$-point functions at $z=0$. The failure occurs at 
higher wavenumbers at $z=2$. In the context of
SPT, this may signal a breakdown of the perfect fluid description of
the dark matter density field at and beyond these wavenumbers. In other
words, even if computed to all orders, SPT (and its variants such as RPT
\cite{rpt}) fails to describe the nonlinear structure formation
beyond these wavenumbers. Therefore, our results yields a quantitative
estimate for the scales at which effective fluid corrections become
important in the bispectrum and higher $N$-point functions, and at which 
one should stop trusting pure SPT calculations.

Finally, we point out that the approach presented here can be augmented to measure
more general squeezed-limit $N$-point functions, by including the
response to long-wavelength tidal fields and by considering the
response of small-scale $n$-point functions in addition to the
small-scale power spectrum considered here.

\acknowledgments
We like to thank Liang Dai, Wayne Hu, Yin Li, Uro$\check{\rm s}$ Seljak, Masahiro Takada, and Matias Zaldarriaga for helpful discussions.

%%%%%%%%%%%%%%%%%%%%%%%%%%%%%%%%%%%%%%%%%%%%%%%%%%%%%%%%%%%%%%%%%%%%%%%%%%%%
%%%%%%%%%%%%%%%%%%%%%%%%%%%%%%%%%%%%%%%%%%%%%%%%%%%%%%%%%%%%%%%%%%%%%%%%%%%%
\appendix

%%%%%%%%%%%%%%%%%%%%%%%%%%%%%%%%%%%%%%%%%%%%%%%%%%%%%%%%%%%%%%%%%%%%%%%%%%%
\section{Squeezed limit $N$-point functions and power spectrum response}
\label{app:sqlimit}

In this appendix, we prove the relation \refeq{Rnpoint} between the
power spectrum response and the squeezed limit $N$-point functions.  
We only consider equal-time $N$-point functions, to which there are no
boost-type contributions from kinematical consistency relations.  
Further, we assume that the long-wavelength modes are well inside the
horizon, removing gauge-dependent terms present for horizon-scale
modes.  Note that the relations derived here retain their formal
validity even for horizon-scale long-wavelength modes if the 
matter density perturbation is evaluated in synchronous-comoving
gauge \cite{CFCpaper2}.  

As in \refeq{Pkexpansion}, we expand the power spectrum as a function of the
\emph{linearly extrapolated initial overdensity} $\d_{L0}$ as 
\be
P(k, t|\d_{L0}) = \sum_{n=0}^\infty \frac1{n!} R_n(k,t) \left[\d_{L0} \hat D(t)\right]^n\: P(k, t)\,,
\label{eq:PkexpansionA}
\ee
where $R_n(k,t)$ are response functions with $R_0(k,t) = 1$.  At the same
order in derivatives, that is at the same order in $k_i/k$ of the 
squeezed-limit $N$-point function, the power spectrum will also depend
on the long-wavelength tidal field which can be parametrized through
\be
K_{ij}(\vk) \equiv \left(\frac{k_i k_j}{k^2} - \frac13 \delta_{ij}\right) \d(\vk)\,.
\ee
Exploring the response of the power spectrum to a long-wavelength tidal
field is beyond the scope of this paper.  We remove the dependence on
$K_{ij}$ by performing an angle-average of the long-wavelength modes which
cancels the tidal field contributions.  

In the following,
we will suppress the time argument for clarity.  We let $\mathcal{S}_n$
be defined as in \refeq{Sdef},
\ba
\mathcal{S}_n(k, k'; k_1,\cdots,k_n) \equiv\:& \int \frac{d^2\hat{\vk}_1}{4\pi} \cdots
\int \frac{d^2\hat{\vk}_n}{4\pi}
\< \d(\vk) \d(\vk') \d(\vk_1) \cdots \d(\vk_n) \>'_c \,.
%\label{eq:Sdef}
\ea
Here, the prime denotes that the factor $(2\pi)^3 \d_D(\vk+\vk'+\vk_{1\cdots n})$
is dropped, where $\vk_{1\cdots n} = \sum_{i=1}^n \vk_i$.   
We consider the limit
\be
\lim_{k_i\to 0}\; \frac{\mathcal{S}_n(k,k'; k_1,\cdots k_n)}{P(k) P_l(k_1) \cdots P_l(k_n)}\,,
\ee
which means that \emph{all} $|\vk_i|$ are taken to zero.   
In this limit, spatial homogeneity enforces $\vk' = -\vk + \O(k_i/k)$, so
that (for statistically isotropic initial conditions) the r.h.s. only
depends on $k$.  

In order to prove \refeq{Rnpoint}, we first note that since we are 
interested in the limit $k_i \to 0$, we can replace $\d(\vk_i)$ in
\refeq{Sdef} with the linear density field $\d_L(\vk_i)$.  We further
transform $\vk_i$ into real space, writing
\ba
\mathcal{S}_n(k; k_1,\cdots,k_n) =\:& \int \frac{d^2\hat{\vk}_1}{4\pi} \cdots
\int \frac{d^2\hat{\vk}_n}{4\pi}
\prod_{i=1}^n
\int d^3 \vx_i\: e^{i \vx_i \cdot \vk_i}
\< \d(\vk) \d(\vk') \d(\vx_1) \cdots \d(\vx_n) \>'_c \vs
=\:& \prod_{i=1}^n \int d^3 \vx_i\: e^{i \vx_i \cdot \vk_i}\:\tilde{\mathcal{S}}_n(k, x_1, \cdots x_n)\,,
\ea
where 
\be
\tilde{\mathcal{S}}_n(k;x_1,\cdots x_n) \equiv \int \frac{d^2\hat{\vx}_1}{4\pi}
\cdots \int \frac{d^2\hat{\vx}_n}{4\pi}
 \< \d(\vk) \d(\vk') \d(\vx_1) \cdots \d(\vx_n) \>'_c\,,
\label{eq:Stdef}
\ee
Note that the angle average is a linear operation and thus commutes with
the Fourier transform; in other words, the $k$-space angle average of the 
Fourier transform of a function is the Fourier transform of the $x$-space 
angle average of the same function.  

Now consider the limit $k_i \to 0$, which implies that $x_i \to \infty$
in the argument of $\tilde{\mathcal{S}}_n$.  Then $\tilde{\mathcal{S}}_n(k)$ 
describes the modulation of the small-scale power spectrum $P(k, \v{0})$ 
measured around $\vx=0$ by 
$n$ spherically symmetric large-scale modes (recall that $\vk' \approx -\vk$).  
This statement can be formalized by introducing
an intermediate scale $R_L$ such that $1/k \ll R_L \ll |\vx_i| \sim 1/k_i$ and
defining $\d(\vk) \to \d_{R_L}(\vk)$ to be the Fourier transform within a cubic 
volume of size $R_L$ around $\vx=0$.  
Then, $\d_{R_L}(\vk) = \d(\vk) + \O(1/(kR_L))$,
while the long-wavelength modes are constant over the same volume with
corrections suppressed by $k_i R_L$.  The corrections we expect in the end
are thus of order $k_i/k$.  
To lowest order in these corrections, $\tilde{\mathcal{S}}_n(k)$ can
be written as
\ba
\lim_{k_i\to 0}:\quad
\tilde{\mathcal{S}}_n(k;x_1,\cdots x_n) =\:& \int \frac{d^2\hat{\vx}_1}{4\pi}
\cdots \int \frac{d^2\hat{\vx}_n}{4\pi}
 \< P(k, \v{0}) \d_L(\vx_1) \cdots \d_L(\vx_n) \>'_c  \,.
\ea
We can now insert the expression for the local power spectrum from 
\refeq{PkexpansionA}, which immediately yields
\ba
\lim_{k_i\to 0}:\quad\tilde{\mathcal{S}}_n(k;x_1,\cdots x_n) =\:&
\sum_{m=0}^\infty \frac1{m!} R_m(k) P(k) 
\int \frac{d^2\hat{\vx}_1}{4\pi}
\cdots \int \frac{d^2\hat{\vx}_n}{4\pi}
\< \d_L^m(\v{0}) \d_L(\vx_1) \cdots \d_L(\vx_n) \>'_{i-0} \,.
\label{eq:rspace2}
\ea
Here, the subscript $i-0$ indicates that only contractions between $\v{0}$
and $\vx_i$ are to be taken, since the l.h.s. is defined through the
connected correlation function (all other contractions would contribute to 
the disconnected part of  $\< \d(\vk) \d(\vk') \d(\vx_1) \cdots \d(\vx_n) \>$).  
Since all density fields in the correlator in \refeq{rspace2} are linear,
limiting to the contractions between $\v{0}$ and $\vx_i$ then constrains 
$m=n$.\footnote{If we use the definition of Eulerian response in \refeq{Pkexpansion} instead, we 
equivalently obtain a slightly more complicated relation where all $m \leq n$ contribute.}  
We obtain
\ba
\lim_{k_i\to 0}:\quad \tilde{\mathcal{S}}_n(k;x_1,\cdots x_n) =\:& \frac1{n!} R_n(k) P(k)\: n! 
\prod_{i=1}^n  \int \frac{d^2\hat{\vx}_i}{4\pi}
\:\xi_L(\vx_i) \vs
=\:&  R_n(k) P(k) \prod_{i=1}^n
\int \frac{d^3 \vk_i}{(2\pi)^3} e^{i \vx_i \cdot \vk_i} P_L(k_i)\,.
\ea
Here, $\xi_L$ and $P_L$ denote the linear matter correlation function and
power spectrum, respectively.  The angle average in the first line
is trivial of course.  Going back to Fourier space then immediately yields
that for $k_i \to 0$,
\be
\mathcal{S}_n(k; k_1,\cdots k_n) = 
R_n(k) P(k) \prod_{i=1}^n P_L(k_i)
+ \O\left(\frac{k_i}{k},\:\frac{k_i}{k_{\rm NL}}\right) \,,
\ee
where $k_{\rm NL}$ is the nonlinear scale.  This can be reordered to yield \refeq{Rnpoint},
\be
R_n(k) = \lim_{k_i\to 0}\; \frac{\mathcal{S}_n(k; k_1,\cdots k_n)}{P(k) P_L(k_1) \cdots P_L(k_n)}\,.
\nonumber
\ee

This provides the connection between the response functions $R_n(k)$ and
the angle-averaged matter $(n+2)$-point function \refeq{Sdef} in a certain
limit squeezed limit (since $k_i \ll k$).  Note that
no assumption about the magnitude of $k$ has been made, i.e. this value
can be fully nonlinear.  In this paper, we accurately determine this fully
nonlinear quantity using simulations.  In the following subsections, we
illustrate \refeq{Rnpoint} at tree level in perturbation theory for the
cases $n=1$ (three-point function) and $n=2$ (four-point function).

\subsection{Tree-level result: $n=1$}

At tree-level for $n=1$ we obtain
\ba
\lim_{k_i \to 0} \mathcal{S}_1(k; k_1) \stackrel{\rm tree-level}{=} \:&
2 \lim_{k_1\to 0}\:\int \frac{d^2\hat{\vk}_1}{4\pi}
\left[F_{2}(\vk, \vk_1) P_l(k) + F_{2}(-\vk-\vk_{1}, \vk_1) P_l(|\vk+\vk_{1}|)\right] P_l(k_1) \,.
\ea
\refeq{Rnpoint} then yields
\be
R_1(k) \stackrel{\rm tree-level}{=} \frac{2}{P_l(k)}
\lim_{k_1\to 0}\:\int \frac{d^2\hat{\vk}_1}{4\pi}
\left[F_{2}(\vk, \vk_1) P_l(k) + F_2(-\vk-\vk_{1}, \vk_1) P_l(|\vk+\vk_{1}|)\right] \,.
\label{eq:R1F2}
\ee
where 
\be
F_2(\vk, \vk_1) = \frac57 + \frac12 \mu \left(\frac{k}{k_1} + \frac{k_1}k\right) + \frac27 \mu^2\,,
\ee
where $\mu$ is the cosine of $\vk$ and $\vk_1$. The term $\mu/2 (k/k_1)$
is problematic as we are sending $k_1 \to 0$.  Using that
\ba
|\vk+\vk_1| =\:& k ( 1 + q \mu + \O(q^2)), \quad q = \frac{k_1}k \vs
P_l(|\vk+\vk_1|) =\:& P_l(k) \left[1 + \frac{d\ln P_l(k)}{d\ln k} q \mu + \O(q^2)\right]\,,
\ea 
the sum of the two IR-divergent terms in \refeq{R1F2} becomes
\ba
\frac12 \left\{ \mu \frac{k}{k_1} P_l(k) + \frac{-(\vk+\vk_1)\cdot \vk_1}{|\vk+\vk_1| k_1} P_l(|\vk+\vk_1|) \right\} =
\frac12 P_l(k) \left[ - \mu^2 \frac{d\ln P_l(k)}{d\ln k} - 1\right] + \O(q)\,.
\ea
As expected, the divergent pieces have canceled.  We have dropped terms
of order $k_1/k$ which are irrelevant in the limit we are interested in.  
We finally obtain
\ba
R_1(k) \stackrel{\rm tree-level}{=}\:& 2 \int_{-1}^1 \frac{d\mu}2
\left[\frac{10}7 - \frac12 \left(\mu^2 \frac{d\ln P_l}{d\ln k} + 1 \right)
+ \frac47 \mu^2 \right] \vs
=\:& 2\left[\frac{10}7 -\frac16 \frac{d\ln P_l}{d\ln k} - \frac12 + \frac4{21}\right] = \frac{47}{21} - \frac13 \frac{d\ln P_l}{d\ln k}\,.
\label{eq:R1F2r}
\ea
This agrees with the linear prediction for $R_1$, obtained from substituting
$f_1$, $e_1$, and $g_1$ into \refeq{R1}.

\subsection{Tree-level result: $n=2$}

At $n=2$, we have
\ba
\mathcal{S}_n(k; k_1, k_2) =\:& \int \frac{d^2\hat{\vk}_1}{4\pi}
\int \frac{d^2\hat{\vk}_2}{4\pi}
\< \d(\vk) \d(\vk') \d(\vk_1) \d(\vk_2) \>'_c \vs
\stackrel{k_1,k_2\to 0}{=}\:& \int \frac{d^2\hat{\vk}_1}{4\pi} \int \frac{d^2\hat{\vk}_2}{4\pi} \Bigg\{  6 \Big[ F_3(\vk, \vk_1, \vk_2) P_l(k) \vs
&+ F_3(-\vk-\vk_{12}, \vk_1, \vk_2) P_l(|\vk+\vk_{12}|)\Big] \vs
& + 4 \Big[ F_2(-\vk_1, \vk+\vk_1) F_2(\vk_2, \vk+\vk_1) P_l(|\vk+\vk_1|)\vs
 & + F_2(\vk_1, \vk+\vk_2) F_2(-\vk_2,\vk+\vk_2) P_l(|\vk+\vk_2|) \Big] \Bigg\}P_l(k_1) P_l(k_2)
\,,
\ea
where $F_3$ is the symmetrized third-order perturbation theory kernel, and the unsymmetrized form can be found
in \cite{goroff/etal:1986}.  Both $F_2^2$ and $F_3$ contain formally IR-divergent
terms up to $\O[(q_{1,2})^2]$ where $q_{1,2}=k_{1,2}/k$ for $k_{1,2} \to 0$ which
cancel in the end, so
we should expand the power spectrum to $\O[(k_{1,2})^2]$ to obtain the consistent
result at order $\O[(q_{1,2})^0]$.  We have
\ba
P_l(|\vk+\vk_1|) =\:& P_l(k) \left[1 + \left(q_1\mu_1+\frac{q_1^2}{2}[1-\mu_1^2]\right) \frac{k}{P_l(k)} \frac{dP_l(k)}{dk}
+ \frac{q_1^2\mu_1^2}{2} \frac{k^2}{P_l(k)} \frac{d^2P_l(k)}{dk^2} + \O(q_1^3)\right] \vs
P_l(|\vk+\vk_1+\vk_2|) =\:& P_l(k) \Bigg[1 + \Big( [q_1\mu_1+q_2\mu_2] + \frac{1}{2}
\big[q_1^2 (1-\mu_1^2) + q_2^2 (1-\mu_2^2) \vs 
& + 2 q_1 q_2 (\mu_{12}-\mu_1\mu_2)\big]\Big)
\frac{k}{P_l(k)} \frac{dP_l(k)}{d\ln k} \vs
\:& + \frac{1}{2}\left(q_1^2\mu_1^2+q_2^2\mu_2^2+2q_1q_2\mu_1\mu_2\right)
\frac{k^2}{P_l(k)} \frac{d^2P_l(k)}{dk^2} + \O[(q_{1,2})^3]\Bigg]
\,,
\ea
where $\mu_{1,2}$ is the cosine of $\vk$ and $\vk_{1,2}$, and $\mu_{12}$ is the
cosine of $\vk_1$ and $\vk_2$. The leading order terms are
\ba
R_2(k) \stackrel{\rm tree-level}{=}\:& \int \frac{d^2\hat{\vk}_1}{4\pi}
\int \frac{d^2\hat{\vk}_2}{4\pi} \frac{1}{147} \Bigg[ (628+324\mu_1^2+112\mu_{12}^2 \vs 
& -280\mu_1\mu_2\mu_{12}+380\mu_2^2 +656\mu_1^2\mu_2^2-56\mu_2^2\mu_{12}^2) \vs
\:& + (-273\mu_1^2+147\mu_1\mu_2\mu_{12}-336\mu_2^2-483\mu_1^2\mu_2^2+63\mu_2^2\mu_{12}^2) \frac{k}{P_l(k)} \frac{dP_l(k)}{dk} \vs
\:& + 147\mu_1^2\mu_2^2 \frac{k^2}{P_l(k)} \frac{d^2P_l(k)}{dk^2} \Bigg] \vs
=\:& \frac{8420}{1323}-\frac{100}{63}\frac{k}{P_l(k)} \frac{dP_l(k)}{dk}
+\frac{1}{9} \frac{k^2}{P_l(k)} \frac{d^2P_l(k)}{dk^2}
\,,
\label{eq:R2F2F3r}
\ea
which is in agreement with the linear prediction of \refeq{R2} once
the numbers for $e_i,\,f_i,\,g_i$ $(i=1,2)$ are inserted.

%%%%%%%%%%%%%%%%%%%%%%%%%%%%%%%%%%%%%%%%%%%%%%%%%%%%%%%%%%%%%%%%%%%%%%%%%%%
\section{Analytical solution for $\d_\rho$ and $\delta_a$ in Einstein-de Sitter}
\label{app:clMD}

As derived in \cite{sep1}, the modified cosmology described by $\tilde a(t)$
follows the evolution equation of a uniform density spherical perturbation
in the fiducial cosmology.  In this section, we take the fiducial cosmology
to be Einstein-de Sitter (EdS), $\Om = 1$, and consider a positive overdensity
perturbation.  In this case there is a well known
parametric solution for the collapse \cite{peebles:1974}:
\ba
\tilde a(\theta) =\:& \frac12 \frac{\tOm}{\tOm-1} (1-\cos\theta) \vs
\tilde H_0 t(\theta) =\:& \frac12 \frac{\tOm}{(\tOm-1)^{3/2}} (\theta - \sin\theta)\,.
\label{eq:psol}
\ea
Note that $\tOm - 1 = -\tOK > 0$.  We now want to derive the matching
between $\tilde H_0$ and $\tOm$ with the fiducial values and the
linearly extrapolated overdensity $\d_{L0}$.  The same matching also
works for $\d_{L0} < 0$ in which case the $\sin,\,\cos$ are to
be replaced by $\sinh,\,\cosh$.  First,
the $\theta\to 0$ limit should reduce to EdS, where
\be
\tilde a(t) = \left(\frac32 \tilde H_0 t\right)^{2/3}\,,
\label{eq:aEdS}
\ee
so that $\tilde a(t_0) = 1$ and $\tilde H(t_0) = \dot{\tilde{a}}(t_0)/\tilde a(t_0) = \tilde H_0$. 
One can further expand \refeq{psol} around $\theta =0$.   
On the other hand, we know the linear relation between $\tilde a(t)$ 
and $a(t)$, since at linear order $\d_a = -\d_\rho/3 = - \d_{L0} a(t)/3$:
\ba
\tilde a(t) = a(t) [1 + \d_a(t) ] = \left(\frac32 H_0 t\right)^{2/3}
\left[1 - \frac13 \d_{L0} \left(\frac32 H_0 t\right)^{2/3} \right]\,,
\ea
where we have used that $a(t_0)=1$ and $\d_{L0}$ is the linearly extrapolated
overdensity at $t_0$.  Matching at zeroth order in $(H_0 t)^{2/3}$ yields
an identity.  At linear order in $\d_{L0}$, we obtain
\ba
\frac{\tOm -1}{\tOm} =\:& \frac53 \d_{L0} 
\quad\mbox{and}\quad
1 - (1+\d_H)^2 = \frac53 \d_{L0}\,,
\ea
where $1+\d_H = \tilde H_0/H_0$.  
Thus, we have determined all quantities in \refeq{psol} in terms of
$\d_{L0}$.  

One might wonder what happens in the limit $\d_{L0} \to 3/5$, 
which implies $\tilde K/H_0^2 \to 1$, 
$\tOm \to \infty$ and $\d_H \to -1$ and thus seemingly an ill-defined
cosmology.  Let us investigate the solution
\refeq{psol} in this limit.  First, we have
\ba
H_0 t(\theta) =\:& \frac12 \frac{(1+\d_H)^{-3}}{(\tOm-1)^{3/2}} (\theta - \sin\theta) =
\frac12 \left(1 - (1+\d_H)^2\right)^{-3/2} (\theta - \sin\theta)
\,.\vs
\ea
For $\d_H\to -1$ and $\tOm\to\infty$, this solution becomes
\ba
\tilde a(\theta) =\:& \frac12 (1-\cos\theta) \vs
H_0 t(\theta) =\:& \frac12 (\theta-\sin\theta)\,.
\label{eq:psollimit}
\ea
Thus, the solution remains perfectly valid.  It is simply the parametrization
in terms of $\tilde H_0$ and $\tOm$ which breaks down.  We also see why
this happens: turn-around happens in \refeq{psollimit} at $\theta=\pi$,
$H_0 t_{\rm ta} = \pi/2$, and $a_{\rm ta} = 1$.  Thus, the Hubble constant
goes to zero just at the point where we are trying to match the Friedmann
equation, which of course assumes an expanding universe.  
For even larger overdensities, $a(\theta)$ never reaches unity and
thus no matching to $\tilde H_0,\:\tOm$ is possible.  Note that solutions
of course exist, they are simply not captured by a parametrization of
the form \refeq{psol}.  In any case, such large nonperturbative values of the long-wavelength
overdensity are not of practical interest for the application to separate universe
simulations.

%%%%%%%%%%%%%%%%%%%%%%%%%%%%%%%%%%%%%%%%%%%%%%%%%%%%%%%%%%%%%%%%%%%%%%%%%%%
\subsection{Perturbative solution and nonlinear growth factor}
\label{app:growthMD}

In this section we derive the series solution \refeq{seriessol} for
$\d_a$ and $\d_\rho$ in EdS.  
We write the parametric solution as 
\ba
\tilde a(\theta) =\:& \frac12  \eps^{-1} (1-\cos\theta) \vs
 \hat t(\theta) \equiv \frac{t(\theta)}{t_0} 
%=\:& \frac32\:\frac12 \eps\frac{\tOm}{(\tOm-1)^{3/2} (1+\d_H)} (\theta - \sin\theta) \vs
%
=\:& \frac34 \eps^{-3/2} (\theta - \sin\theta) \,,
\ea
where we have used $a(t) = (t/t_0)^{2/3}$, $t_0 = 2/(3H_0)$, and defined
\be
\eps \equiv \frac{\tOm-1}{\tOm} 
%= \left(1-\frac53\d_{L0}\right) 
%\left[ \left(1-\frac53\d_{L0}\right)^{-1} - 1\right] 
= \frac53 \d_{L0}\,.
\label{eq:epsdef}
\ee
Our goal is to obtain 
\be
\tilde a(t_0) = 1 + \d_a(t_0)\,.
\ee
Thus, we need to solve
\be
1 = \hat t(\theta_0) \quad\Leftrightarrow\quad
\frac43 \eps^{3/2} = \theta_0 - \sin\theta_0
\label{eq:theta0eq}
\ee
for $\theta_0$.  We perform a series expansion,
\be
\theta_0 - \sin\theta_0 = \frac16 \theta_0^3 - \frac1{120} \theta_0^5 + \cdots = \sum_{n=1}^\infty b_n \theta_0^{2n+1}\,,
\ee
and solve \refeq{theta0eq} order by order.  The leading solution is 
$\theta_0^{(1)} = 2 \eps^{1/2}$.  The $n$-th order solution has to solve
\be
\frac43 \eps^{3/2} = \sum_{k=1}^n b_k \left[\theta_0^{(n-k+1)}\right]^{2k+1}\,.
\ee
Note that in order to trust the final expression at order $\delta_{L0}^m$, this
solution needs to be expanded to order $n=m+2$.  In the following, we choose
$m=5$.  Solving this order by order, we obtain
\be
\theta_0 = 2 \eps^{1/2} \left[1 + \frac1{15} \eps + \frac2{175} \eps^2 + \frac4{1575} \eps^3 + \frac{43}{67375} \eps^4 + \cdots  \right]\,.
\ee
We then insert this into $\tilde a(\theta)$ and expand in $\eps$, replacing
$\eps$ with $5/3\:\d_{L0}$ through \refeq{epsdef}.  This yields
\be
1 + \d_a(t_0) = \tilde a(\theta_0) = 1 + \sum_{n=1}^\infty e_n \d_{L0}^n \,,
\ee
where the first few coefficients are
\be
e_1 = - \frac13;\quad
e_2 = - \frac1{21}; \quad
e_3 = - \frac{23}{1701}; \quad
e_4 = - \frac{1894}{392931}; \quad
e_5 = - \frac{3293}{1702701}\,.
\label{eq:en}
\ee
We can generalize this to other times $t$ by replacing $\d_{L0}$ with
$\d_{L0} a(t)$, which is possible since EdS is scale free:
\be
\d_a(t) = \sum_{n=1}^\infty e_n [\d_{L0} a(t)]^n \,.
\label{eq:dapert}
\ee
Finally, we obtain the density at $t$ through
\ba
\d_\rho(t) = [1 + \d_a(t)]^{-3} - 1\,,
\ea
and expanding in powers of $\d_{L0}$.  This yields
\ba
\d_\rho(t) =\:& \sum_{n=1}^\infty f_n [\d_{L0} a(t)]^n 
\vs
f_1 =\:& 1; \quad
f_2 = \frac{17}{21}; \quad
f_3 = \frac{341}{567}; \quad
f_4 = \frac{55805}{130977};\quad
f_5 = \frac{213662}{729729}\,.
\label{eq:fn}
\ea
We have verified that \refeq{dapert} and \refeq{fn} are accurate in 
$\Lambda$CDM as well when replacing $a(t)$ with $D(t)/D(t_0)$, where
$D(t)$ is the growth factor in the fiducial cosmology. 

%%%%%%%%%%%%%%%%%%%%%%%%%%%%%%%%%%%%%%%%%%%%%%%%%%%%%%%%%%%%%%%%%%%%%%%%%%%
\section{Linear growth in modified cosmology}
\label{app:smallgrowth}

In this section we iteratively solve the growth factor $\tilde D(t)$ in the modified
cosmology, for a fiducial EdS background, to obtain a
perturbative expansion in terms of $\d_{L0}$ [\refeq{Dtilde}].  
The modified cosmology is described by
\ba
\tilde H^2(\tilde a) =\:& \tilde H_0^2 \left[ \tOm \tilde a^{-3} + (1-\tOm) \tilde a^{-2} \right] \vs
=\:& H_0^2 \left[ \tilde a^{-3} - \eps \tilde a^{-2} \right]\,,
\ea
where $\eps$ is defined in \refeq{epsdef} and we have used $\tOm \tilde H_0^2 = \Om H_0^2 = H_0^2$.  Similarly, the mean background density in the curved
universe is given by
\be
4\pi G \tilde{\rhob}(t) = \frac32 \tOm \tilde H_0^2 \tilde a^{-3}(t) = 
\frac32 H_0^2 \tilde a^{-3}(t)\,.
\ee
On the other hand, in the background EdS cosmology we have
\be
H(a) = H_0 a^{-3/2}\,.
\ee
We want to derive the growth of density perturbations 
$\tilde\d_s = \rho/\tilde{\rhob}-1$, where the subscript $s$ denotes that
they are small-scale fluctuations, in order to distinguish from the
density perturbations $\d_\rho = \tilde{\rhob}/\rhob-1$ with respect to 
the EdS background.  Note also that $\delta_s$ is defined with respect
to the background density of the curved universe (i.e. we do not include the ``reference density'' contribution here).  The growth equation is then given by
\be
\ddot{\tilde\d}_s + 2 \tilde H(t) \dot{\tilde\d}_s - 4\pi G\tilde{\rhob}(t) \tilde\d_s = 0\,,
\ee
which becomes
\ba
\ddot{\tilde\d}_s + 2 H_0 (\tilde a^{-3} - \eps \tilde a^{-2})^{1/2} \dot{\tilde\d}_s
- \frac32 H_0^2 \tilde a^{-3} \tilde\d_s =\:& 0 \vs
\ddot{\tilde\d}_s + 2 H_0 \tilde a^{-3/2} (1 - \eps \tilde a)^{1/2} \dot{\tilde\d}_s
- \frac32 H_0^2 \tilde a^{-3} \tilde\d_s =\:& 0
\,.
\ea
Note that we have neglected the curvature contribution to the Poisson
equation, which involves $3 K \Phi$ and the curved-space Laplacian.  
If $\tilde K/H_0^2 \sim 1$, these terms only become relevant for small-scale modes
that are of order the horizon.  Since we are studying the subhorizon evolution
of small-scale modes, and moreover we always have $\tilde K/H_0^2 \ll 1$ for
a flat fiducial cosmology, these terms are entirely negligible for our purposes.  
We now replace $t$ with $y = \ln a(t)$ as time coordinate, where $a(t)$ is
the scale factor in the EdS background.  
%We have
%\be
%\frac{d}{dt} = H\frac{d}{dy}\:;\quad
%\frac{d^2}{dt^2} = H^2 \left(\frac{d^2}{dy^2} - \frac32 \frac{d}{dy} \right)\,,
%\ee
%where $H(t) = H_0 a^{-3/2}(t)$.  
Dividing by $H^2$, and inserting $\eps = 5/3 \d_{L0}$, we obtain
\be
\frac{d^2}{dy^2} \tilde\d_s + \left[2 \D_a^{-3/2} \left(1 - \frac53 \d_{L0} a \D_a \right)^{1/2} - \frac32 \right] \frac{d}{dy} \tilde\d_s - \frac32 \D_a^{-3} \tilde\d_s = 0\,,
\label{eq:tdseq}
\ee
where we have defined $\D_a(t) \equiv 1 + \d_a(t)$.  So far, everything is
exact.  We now perform a series expansion of \refeq{tdseq} in $\d_{L0}$.  
At zeroth order, $\D_a = 1$, and we obtain
\be
\frac{d^2}{dy^2} \tilde\d_s^{(0)} + \frac12 \frac{d}{dy} \tilde\d_s^{(0)} 
- \frac32 \tilde\d_s^{(0)} = 0\,,
\ee
which has growing and decaying modes of $\tilde\d_s^{(0)} \propto e^y$ and
$e^{-3y/2}$.  Thus, it is identical to the growth in the background EdS
cosmology, as expected.  In the following, we will drop the decaying mode
following standard practice.  Further, we will normalize $\tilde\d_s^{(0)}$
to $a(t) = e^y$ at early times, and replace it with $\tilde D$ to denote the
small-scale growth factor.  Thus, $\tilde D^{(0)}(t) = a(t)$.  

The series expansion of $\D_a$ is given in \refeq{dapert}, which we
can generalize to other times $t$ by replacing $\d_{L0}$ with $\d_{L0} a(t)$.  
We see that \refeq{tdseq} can be expanded into a series in powers of
$\d_{L0} a$, leading to
\be
\frac{d^2}{dy^2} \tilde D 
+ \left[ \sum_{m=0}^\infty c_m \d_{L0}^m e^{m y} \right] \frac{d}{dy} \tilde D 
- \left[ \sum_{m=0}^\infty d_m \d_{L0}^m e^{m y} \right] \tilde D = 0\,,
\ee
with coefficients $c_m,\:d_m$.  Specifically, the coefficients are defined
through
\ba
2 \D_a^{-3/2} \left(1 - \frac53 \d_{L0} a \D_a \right)^{1/2} - \frac32
=\:& \sum_{m=0}^\infty c_m [\d_{L0} a(t)]^m \vs
\frac32 \D_a^{-3} =\:& \sum_{m=0}^\infty d_m [\d_{L0} a(t)]^m \,.
\ea
Correspondingly, we write the pure growing-mode solution as a series
\be
\tilde D(y) = \sum_{n=0}^\infty g_n \d_{L0}^n e^{(n+1) y}\,,
\ee
with coefficients $g_n$.  Given our choice of normalization, we have $g_0 = 1$.  Thus,
\be
\frac{d}{dy}\tilde D(y) = \sum_{n=0}^\infty (n+1) g_n \d_{L0}^n e^{(n+1) y}; \quad
\frac{d^2}{dy^2} \tilde D(y) = \sum_{n=0}^\infty (n+1)^2 g_n \d_{L0}^n e^{(n+1) y}\,.
\ee
Assuming we have a solution to order $n-1$, the solution
at order $n$ then has to satisfy
\be
(n+1)^2 g_n \d_{L0}^n e^{(n+1)y} + \sum_{m=0}^n g_{n-m} \d_{L0}^{n-m} e^{(n-m+1)y} \left[(n-m+1) c_m \d_{L0}^m e^{m y} - d_m \d_{L0}^m e^{m y} \right] = 0\,.
\ee
The time dependence $e^{n y}$ factors out, and we obtain a simple algebraic
relation for $g_n$ in terms of $\{c_m,\:d_m,\:g_m\}_{m=0}^{n-1}$:
\be
(n+1)^2 g_n + \sum_{m=0}^n g_{n-m} \left[ (n-m+1) c_m - d_m \right] = 0\,.
\ee
For example, for $n=1$ we have
\be
4 g_1 + \left[2 c_0 - d_0\right] g_1 + g_0 \left[c_1 - d_1\right] = 0\,,
\ee
where $c_0 = 1/2,\:d_0 = 3/2,\:c_1 = -2/3,\:d_1 = 3/2$.  Thus,
\be
\frac72 g_1 = \frac23 + \frac32 = \frac{13}{6}
\quad\Rightarrow\quad
g_1 = \frac{13}{21}\,.
\ee
This is the result for the leading $\O(\d_{L0})$ 
correction to the small-scale growth.  Straightforward algebra yields
the extension to higher order.  We thus obtain
\be
\tilde D(t) = a(t) \left[1 + \sum_{n=1}^\infty g_n [ \d_{L0} a(t) ]^n \right]\,,
\ee
where the first few coefficients are
\ba
g_1 = \frac{13}{21};\quad
g_2 = \frac{71}{189}; \quad
g_3 = \frac{29609}{130977}; \quad
g_4 = \frac{691858}{5108103}\,.
\label{eq:gn}
\ea
In order to generalize from EdS to other cosmologies, we
perform the usual replacement of $a(t)\to D(t)$, where $D(t)$ is the 
growth factor in the fiducial cosmology normalized to $a(t)$ during
matter domination.  Thus, we obtain
\be
\tilde D(t) = D(t)\left\{1 + \sum_{n=1}^\infty g_n \left[\d_{L0} \frac{D(t)}{D(t_0)}\right]^n \right\}\,.
\label{eq:smallgrowth}
\ee

%%%%%%%%%%%%%%%%%%%%%%%%%%%%%%%%%%%%%%%%%%%%%%%%%%%%%%%%%%%%%%%%%%%%%%%%%%%
\section{Transformation of power spectrum}
\label{app:xrescalePk}

This section briefly derives the transformation of the power spectrum
under a rescaling of spatial coordinates
\be
\hat{\vx} = c\, \vx\,,
\ee
where $c$ is a constant.  This is necessary in order to calculate the
full power spectrum response, since wavenumbers are defined in comoving
coordinates and the modified scale factor $\tilde a \neq a$ at fixed time.  

The correlation function of a scalar $\d$,
which satisfies $\hat\d(\hat{\vx}) = \d(\vx(\hat{\vx}))$, then transforms
as (App.~A in \cite{conformalfermi})
\be
\hat\xi(\hat{\vr}) = \< \hat\d\left(\frac{\hat{\vr}}{2}\right)
\hat\d\left(-\frac{\hat{\vr}}{2}\right) \>
= \< \d\left(\frac{\hat{\vr}}{2 c}\right)
\d\left(-\frac{\hat{\vr}}{2 c}\right) \> = \xi(c^{-1} \hat{\vr})\,.
\ee
Note that for our purposes $\d$ transforms as scalar, since we take into
account the change in the background reference density separately.  Since
\be
\xi(c^{-1} \hat{\vr}) = \int \frac{d^3\vk}{(2\pi)^3} P(k) \exp\left[ i c^{-1} \vk\cdot\vr\right]\,,
\ee
and $\hat P(\hat{\vk})$ is defined through
\be
\hat\xi(\hat{\vr}) = \int \frac{d^3\hat{\vk}}{(2\pi)^3} \hat P(\hat k) 
\exp\left[ i \hat{\vk}\cdot\hat{\vr}\right]\,,
\ee
one immediately obtains
\be
d^3 \hat{\vk}\:\hat P(\hat{k}) = \left[d^3 \vk\: P(k) \right]_{\vk = c\hat{\vk}}\,.
\ee
Since $c$ is constant, this leads to
\be
\hat P(\hat k) = c^3 P(c\, \hat k).
\label{eq:Prescale}
\ee

%%%%%%%%%%%%%%%%%%%%%%%%%%%%%%%%%%%%%%%%%%%%%%%%%%%%%%%%%%%%%%%%%%%%%%%%%%%%
%%%%%%%%%%%%%%%%%%%%%%%%%%%%%%%%%%%%%%%%%%%%%%%%%%%%%%%%%%%%%%%%%%%%%%%%%%%%
\bibliography{references}
\end{document}